\documentclass[aps,reprint,twocolumn,superscriptaddress,nofootinbib,longbibliography]{revtex4-1}

\usepackage{amsmath, amssymb, bm, braket, dsfont, times, amsthm}
\usepackage{array}
\usepackage{units}
\usepackage{multirow}
\usepackage{graphicx,xfrac,color}
\usepackage[breaklinks,colorlinks,allcolors=blue]{hyperref}
\usepackage{framed}

\definecolor{shadecolor}{gray}{0.95}

\usepackage[normalem]{ulem}

\makeatletter
\def\l@subsection#1#2{}
\def\l@subsubsection#1#2{}

\providecommand{\tabularnewline}{\\}
\makeatother

\usepackage{babel}
\begin{document}
\title{Exceptional entanglement in non-Hermitian fermionic models}
\author{Wei-Zhu Yi}
\affiliation{Shenzhen Key Laboratory of Advanced Quantum Functional Materials
and Devices, Southern University of Science and Technology, Shenzhen
518055, China}
\affiliation{Institute for Quantum Science and Engineering and Department of Physics,
Southern University of Science and Technology, Shenzhen 518055, China}
\author{Yong-Ju Hai}
\affiliation{Institute for Quantum Science and Engineering and Department of Physics,
Southern University of Science and Technology, Shenzhen 518055, China}
\author{Rong Xiao}
\affiliation{National Laboratory of Solid State Microstructures and Department of Physics, 
Nanjing University, Nanjing 210093, China}
\author{Wei-Qiang Chen}
\email{chenwq@sustech.edu.cn}
\affiliation{Shenzhen Key Laboratory of Advanced Quantum Functional Materials
and Devices, Southern University of Science and Technology, Shenzhen
518055, China}
\affiliation{Institute for Quantum Science and Engineering and Department of Physics,
Southern University of Science and Technology, Shenzhen 518055, China}
\begin{abstract}
Exotic singular objects, known as exceptional points, are ubiquitous in non-Hermitian physics. They might be spectral singularities in energy bands that produce anomalous effects and defectiveness. The quantum entanglement of a generic non-Hermitian model with two different types of spectral exceptional points  (SEPs) is systematically investigated in this paper. We discovered a relationship between non-unitary conformal field theories and the $k$-linear-type SEPs, which is typically associated with $\mathcal{PT}$-symmetry or pesdo-Hermicity spontaneous breaking. The underlying association between $k$-square-root-type SEPs, which arise concurrently with real (imaginary) gap closing in the complex spectrum, mimicking first-order-phase-transition criticalities, and complex conformal field theories (cCFTs) is addressed through the calculation of complex central charges. From the entanglement spectrum, zero-energy exceptional modes are found to be distinct from normal zero modes or topological boundary modes. Finally, we include a brief discussion of analogous non-Hermitian quantum spin
models and endeavor to establish an intuitive understanding of exceptional points through the spin picture in various scenarios.

\end{abstract}
\maketitle

\section{Introduction}

Exceptional points (EPs)~\cite{kato_perturbation_2013} are spectral singularities in the parameter space of non-Hermitian quantum systems. At EPs, certain of the eigenvectors and their corresponding eigenvalues coalesce, leading to anomalous degeneracies~\cite{berry_physics_2004,heiss_exceptional_2004}. Due to their rich physics and potential for novel applications, exceptional points and phenomena associated with them have become central subjects of study in non-Hermitian quantum physics~\cite{ashida_non-hermitian_2020,heiss_physics_2012,chan16,non-abelian19,miri_2019,ozdemir_paritytime_2019,yang_fermion_2021,kawabata_symmetry_2019,lai_observation_2019,kawabata_information_2017,dora_kibble-zurek_2019,liao_experimental_2021,kozii_non-hermitian_2017,knots_hu21,knot_yang20,ma20,bergholtz_exceptional_2021,you21}. 
In particular, the topological structures of EPs in dynamically encircling them have been discussed widely~\cite{gao_observation_2015,heiss_circling_2016, ozdemir_paritytime_2019,heiss_collectivity_1998,doppler_dynamically_2016,hassan_dynamically_2017,richter01,li20,milburn15,yu21,liu20} and demonstrated in various systems including optical photonics~\cite{klaiman_visualization_2008,zhen_spawning_2015,cerjan_exceptional_2016,longhi_spectral_2010,zhang14,yang14} and spin, ion, superconducting quantum simulators~\cite{liu2021,ding2021,naghiloo_quantum_2019}.
Other intriguing properties ranging from intrinsic chirality~\cite{ashida_full-counting_2018,Demb2004} to unidirectionality~\cite{regensburger_paritytime_2012,wong_lasing_2016,hassani_gangaraj_topological_2018}, as well as enhanced sensitivity at exceptional points~\cite{chen_exceptional_2017,hodaei_enhanced_2017,zhang_quantum_2019} also attract a lot of attention.

EPs are typically studied in $\mathcal{PT}$-symmetric systems as the gap closing points or the imaginary spectra emerging points due to the spontaneous $\mathcal{PT}$-symmetry breaking. 
Generic EPs, however, will have completely different stories.
While engineering and detecting of exceptional points or their higher-dimensional extensions~\cite{zhen_spawning_2015,xu_weyl_2017,tang_direct_2021} have been accomplished in a variety of few-body systems~\cite{cerjan_exceptional_2016,klaiman_visualization_2008,makris_beam_2008,longhi_spectral_2010,zhang_quantum_2019,chan18}, the features of true many-body systems with EPs, such as their spectral and eigenstate properties, as well as singular behaviors of relevant observables and entanglement properties in the vicinity of EPs, remain largely unexplored.

Gapless spectra emerge along with EPs in the momentum spaces, referred to as spectral exceptional points (SEPs) in bands. In contrast to Hermitian gapless systems, generic non-Hermitian Hamiltonians have complex energy spectrums and the EPs are branch points on the complex energy Riemann surface.

For Hermitian systems, it is well established that the entanglement entropy scales universally as $S \sim (c/3)\log l$ in the $1+1$ dimensional critical system, where $l$ denotes the size of the subsystem and $c$ denotes the central charge of the related conformal field theory~\cite{vidal_entanglement_2003,calabrese_entanglement_2004,Calabrese_2009}.In fermionic systems, $c$ is closely related to the nodal points on the Fermi surface, which cause discontinuities in state occupation~\cite{swingle_entanglement_2010}.
In some gapless non-Hermitian systems without EPs, as in Hermitian systems, the positive central charge is related to the real spectral Fermi points~\cite{herviou_entanglement_2019,guo_entanglement_2021}.

However, when non-Hermitian gapless systems incorporate SEPs, the preceding statement becomes invalid. The central charge has ceased to be positive and no longer displays a naive connection with the Fermi points. Several well-known examples include the 1+1 d Yang-Lee edge singularity~\cite{cardy_conformal_1985,fisher_yang-lee_1978}.
They retain conformal symmetry despite their non-unitary nature and exhibit universal logarithmic scaling of entanglement entropy~\cite{Bianchini_2014,bianchini_entanglement_2015}.
The low-energy theory for such a critical non-Hermitian model is believed to be a non-unitary conformal field theory. 

Recently, non-unitary conformal field theories (nUCFTs) have attracted particular interest. It is demonstrated that nUCFTs not only emerge from the phase transition of classical statistical models or the $\mathcal{PT}$-phase transition of non-Hermitian quantum models, but also are connected to non-unitary topological orders~\cite{lootens_galois_2020,couvreur_entanglement_2017,xu_characterization_2021,bianchini_entanglement_2016,chang_entanglement_2020,chen_galois_2021}.
A major difference between non-unitary and unitary CFTs is that the central charge and conformal weights of some primary fields are negative in non-unitary CFTs. In contrast, they are all positive in unitary CFTs. The physical ground state no longer coincides with the conformal vacuum due to negative conformal weight. Additionally, there is a specialized version of non-unitary conformal field theories with complex central charges known as complex conformal field theories (cCFTs)~\cite{benini_conformality_2020,gorbenko_walking_2018,kaplan_conformality_2009,ma_shadow_2019}.

They might occur in some weakly first-order phase transitions, such as those found in the $Q>4$ Potts model and the $Z_{n}(n>4)$ quantum clock model, as well as in higher-dimensional analogues like confinement-deconfinement quantum phase transitions in quantum magnetism systems. It bears little resemblance to spontaneous symmetry breaking or topological transitions.
Although their entanglement entropy scales logarithmically, whether such quantum criticalities maintain conformal symmetry is still unclear. They lack well-defined universal characteristics: the central charge will flow with the system's size.

In this paper, we study a general non-Hermitian model with exceptional points in the Bloch Hamiltonian using the entanglement measurement. The entanglement spectrum and entropy of the model can be estimated using the modified Peschel's correlation matrix approach described in Sec.~\ref{bee}.
We investigate a two-band non-Hermitian Hamiltonian with tunable non-reciprocal nearest-neighbor hopping and staggered chemical potential in Sec.~\ref{gmodel}. 

In Sec.~\ref{subsec:Linear}, the pseudo-Hermitian model preserving a generalized skew (anti-)$\mathcal{PT}$-symmetry has a linear dispersion around SEPs at the symmetry-breaking point. Due to the presence of SEPs, the correlation matrix eigenvalues and entanglement entropy are unusual. The negative entanglement entropy scales logarithmically with a fitting $c=-2$, unless boundary modes break the conformal (translational) symmetry resulting in a non-universal logarithmic scaling of entanglement entropy. The scaling dimensions of excited states in different boundary conditions provide additional evidence that this is a fermionic ghost CFT. Furthermore, when the electron filling is designed below the EPs, the entanglement entropy reverts to the $c=1$ free fermion instantaneously, emphasizing the significance of the SEPs in determining the entanglement entropy.

In Sec.~\ref{subsec:Linear}, we explore a different type of SEPs, which typically occurs in second-order EPs in the model in Sec.~\ref{gmodel} and happens in EPs in (anti) $\mathcal{PT}$-broken phases. The model with non-reciprocal hopping described in Sec.~\ref{srd} also produces SEPs with a $k$-square-root dispersion. The entanglement entropy scales logarithmically with a complex fitting central charge. However, it is significantly dependent on system size. This, we suggest, implies the existence of a complex conformal field theory underneath. We further present an effective field-theoretical approach to unconventional quantum criticalities by analyzing the flow of the central charge.

The next Sec.~\ref{ESQ} on entanglement spectrum analysis emphasizes the distinction between zero-energy exceptional modes and normal zero modes by introducing a probe bulk mode. This approach also reveals the entanglement features of exceptional bound states.

Finally, in Sec.~\ref{spin}, we analyze several solvable spin-1/2 chain models exhibiting either $\mathcal{PT}$-broken or $\mathcal{PT}$-symmetry spontaneous breaking (SSB) phase transitions and EPs, characterized by either complex drift entanglement entropy or real entanglement entropy associated with confirmed nUCFTs. Moreover, we provide classical spin-based pictures of their respective exceptional points to facilitate understanding.

We believe that our research will contribute to a better understanding of non-Hermitian many-body physics and pave the way for future research into non-unitary field theories and exotic quantum criticalities.

\section{Bi-orthogonal entanglement entropy\label{bee}}
The concept of entanglement entropy can be used to investigate the universal properties of quantum many-body systems. In Hermitian systems, the entanglement entropy can be calculated using the entanglement spectrum derived from the density matrix's singular value decomposition (SVD). However, in non-Hermitian systems, the definitions of density matrix and entanglement entropy are more nuanced. The bi-orthogonal entanglement entropy is used in this paper to characterize the entanglement in non-Hermitian systems.

A proper description of non-Hermitian quantum states is essential, as conventional quantum mechanics fails due to the absence of Hermicity. A prominent case is so-called bi-orthogonal quantum mechanics~\cite{Brody_2013}, which requires two sets of non-self-orthogonal basis sets to completely describe a generic Hamiltonian (Hermitian or not).
Given a (non-Hermitian) Hamiltonian $H$ with eigenvalues
$\varepsilon_{\alpha}$ and corresponding eigenvectors $|\psi_{R}\rangle _{\alpha}$.
Typically, these eigenvectors are not orthogonal, i.e.,
\begin{equation}
    _{\alpha}\langle \psi_{R}\mid\psi_{R}\rangle _{\beta}\neq\delta_{\alpha\beta}.
\end{equation}
However, the eigenvectors of Hermitian conjugate
$H^{\dagger}$, $H^{\dagger}|\psi_{L}\rangle _{\alpha}=\varepsilon_{\alpha}^{*}|\psi_{L}\rangle _{\alpha}$ can be taken into account. The bi-orthogonality states satisfy,
\begin{equation}
    _{\alpha}\langle \psi_{L}\mid\psi_{R}\rangle _{\beta}=\delta_{\alpha\beta}.
\end{equation}
Obviously, when the Hamiltonian is Hermitian, $|\psi_{L}\rangle _{\alpha}=|\psi_{R}\rangle _{\alpha}$,
the norm returns to the ordinary orthogonal relations $_{\alpha}\langle \psi_{R}\mid\psi_{R}\rangle _{\beta}=\delta_{\alpha\beta}$.

With $|\psi_{R}\rangle _{\alpha}$and $|\psi_{L}\rangle _{\alpha}$ as new bases, the origin
Hamiltonian can be diagonalized  as,
\begin{equation}
    H=\sum_{\alpha}\varepsilon_{\alpha}|\psi_{R}\rangle _{\alpha\alpha}\langle \psi_{L}|.
\end{equation}
The expectation values of any operator $\mathcal{O}$ can be evaluated as $\langle\mathcal{O}\rangle=\langle \Psi_{L}|\mathcal{O}|\Psi_{R}\rangle $, where the right state can be expanded with right basis as $|\Psi_{R}\rangle =\sum_{\alpha}c_{\alpha}|\psi_{R}\rangle _{\alpha}$
and the associated left state is $|\Psi_{L}\rangle =\sum_{\alpha}c_{\alpha}|\psi_{L}\rangle _{\alpha}$.
It is worth noting that $\langle \Psi_{L}|\mathcal{O}|\Psi_{R}\rangle =\langle \Psi_{R}|\mathcal{O^{\dagger}}|\Psi_{L}\rangle $ are real observables.

In the language of quantum field theory, the right eigenvector $|\psi_{R}\rangle $
can be constructed by integrating from the present to the infinite
past in the Euclidean space and the left eigenvector $\langle \psi_{L}|$is
the integration from the present to the infinite future~\cite{nishioka}.
The Replica Riemann surface approach is used in this case to determine the entanglement entropy of non-unitary field theories.
In our case of study, the Hamiltonian is a quadratic form. The left and right
eigenstates can be regarded as left and right fermionic creation operators
acting on the vacuum (null state), i.e., $|\psi_{L}\rangle _{\alpha}=\psi_{L,\alpha}^{\dagger}|0\rangle $,
$|\psi_{R}\rangle _{\alpha}=\psi_{R,\alpha}^{\dagger}|0\rangle $.
Remarkably, they respect the fermionic anti-commutation relation,
\begin{equation}
    \{ \psi_{L,\alpha},\psi_{R,\beta}^{\dagger}\} =\delta_{\alpha\beta}.
\end{equation}
With these operators, we can further construct a many-body
steady ``ground'' state, which is the filling-up state according
to the real part of energy levels $\varepsilon_{\alpha}$. Therefore,
we have the left ground state $|GS_{L}\rangle =\prod_{\alpha\in\text{filled}}\psi_{L,\alpha}^{\dagger}|0\rangle $
and right ground state, $|GS_{R}\rangle =\prod_{\alpha\in\text{filled}}\psi_{R,\alpha}^{\dagger}|0\rangle $.
The bi-orthogonal density matrix (or called
right-left density matrix) is then defined as,
\begin{equation}
    \rho_{RL}=|\psi_{R}\rangle \langle \psi_{L}|,
\end{equation} 
which
is natural to describe the steady properties of non-Hermitian systems~\cite{couvreur_entanglement_2017,herviou_entanglement_2019,chang_entanglement_2020}.
Note that $\rho_{RL}^{\dagger}\neq\rho_{RL}$. The reduced density
matrix is calculated by $\rho_{\mathcal{A}}=\text{Tr}{}_{\mathcal{B}}|\psi_{R}\rangle \langle \psi_{L}|$
for a pure state $\rho_{\textrm{tot}}=|\psi_{R}\rangle \langle \psi_{L}|$ consists
of $\mathcal{A}\cup\mathcal{B}$.The (bi-orthogonal) entanglement entropy (BEE) for non-Hermitian systems is then defined
as,
\begin{equation}
    S(\mathcal{A})=-\text{Tr}{}_{\mathcal{A}}\rho_{\mathcal{A}}\ln\rho_{\mathcal{A}}.
\end{equation}
Since the Wick theorem still holds in non-Hermitian cases, we can use Peschel's method to calculate the entanglement entropy through two-point correlation functions~\cite{chung_density-matrix_2001,peschel_calculation_2003,peschel_reduced_2009}.

The correlation function in matrix form is defined as $\mathcal{C}_{ij}=\langle GS_{L}|c_{i}^{\dagger}c_{j}|GS_{R}\rangle $, then the entanglement Hamiltonian $\mathcal{H}_{E}$ is related to the correlation matrix through 
\begin{equation}
\mathcal{C}=\frac{e^{-\mathcal{H}_{E}}}{1+e^{-\mathcal{H}_{E}}},\quad\epsilon_{E}^{\xi}=\ln\left[\left(C_{\mathcal{A}}^{\xi}\right)^{-1}-1\right],
\end{equation}
where $\epsilon_{E}^{\xi}$s are the entanglement Hamiltonian's eigenvalues for subsystem $\mathcal{A}$ and $C_{\mathcal{A}}^{\xi}$s  are the correlation matrix's eigenvalues confined to region $\mathcal{A}$. The bi-orthogonal entanglement entropy can be expressed as,
\begin{equation}
    S(\mathcal{A})=-\sum_{\xi}[C_{\mathcal{A}}^{\xi}\ln C_{\mathcal{A}}^{\xi}+(1-C_{\mathcal{A}}^{\xi})\ln(1-C_{\mathcal{A}}^{\xi})],
\end{equation}
which captures the underlying non-unitary low-energy theories of a particular ground state in a non-Hermitian model.

\section{Non-Hermitian models and quantum entanglement with exceptional points}

\subsection{A generic non-Hermitian model\label{gmodel}}

We explore a non-Hermitian tight-binding model with non-reciprocal nearest-neighbor hopping and staggered chemical potentials.
In real space, the Hamiltonian is given by
\begin{equation}
\label{tmodel}
\begin{split}H= & w_{1}\sum_{n=1}^{L}c_{n+1,A}^{\dagger}c_{n,B}+w_{2}\sum_{n=1}^{L}c_{n,B}^{\dagger}c_{n+1,A}\\
 & +\phantom{}v_{2}\sum_{n=1}^{L}c_{n,B}^{\dagger}c_{n,A}+v_{1}\sum_{n=1}^{L}c_{n,A}^{\dagger}c_{n,B}\\
 & +u\sum_{n=1}^{L}c_{n,A}^{\dagger}c_{n,A}-u\sum_{n=1}^{L}c_{n,B}^{\dagger}c_{n,B}
\end{split}
\end{equation}

where $c_{n,A(B)}^{\dagger}$ and $c_{n,A(B)}$ represents creation
and annihilation operators at the n-th site on A(B) sublattice, $u,v_{1,2},w_{1,2}\in\mathbb{C}$ are
tunable parameters. For simplicity, the parameters take
either purely imaginary or real values. Only when $v_{1}^{*}=v_{2},$ $w_{1}^{*}=w_{2}$
and $u\in\mathbb{R}$, the Hamiltonian returns Hermitian. If we employ
open boundary condition, the famous non-Hermitian skin effect emerges
when $v_{1}^{*}\neq v_{2}$ or $w_{1}^{*}\neq w_{2}$. By imposing
periodic boundary condition (PBC), i.e., $c_{L+1,A(B)}=c_{1,A(B)}$, 
the Hamiltonian in momentum space can be derived via Fourier transformation as
\begin{equation}
\label{hk}
H=\sum_{k}\left(\begin{array}{cc}
c_{k,A}^{\dagger} & c_{k,B}^{\dagger}\end{array}\right)h(k)\left(\begin{array}{c}
c_{k,A}\\
c_{k,B}
\end{array}\right)
\end{equation}
with the single-particle Hamiltonian reads
\begin{equation}
\label{Hk}
h(k)=\left(\begin{array}{cc}
u & w_{1}e^{-ika}+v_{1}\\
w_{2}e^{ika}+v_{2} & -u
\end{array}\right)
\end{equation}
where we take the lattice constant to 1. The dispersion relation is 
\begin{equation}
\label{ek}
\varepsilon_{k}=\pm\sqrt{(a_{r}+b_{r}\cos k)+is\sin k},
\end{equation}
where $a_{r}=w_{1}w_{2}+v_{1}v_{2}+u^{2}$, $b_{r}=w_{2}v_{1}+w_{1}v_{2}$
and $s=w_{2}v_{1}-w_{1}v_{2}$. Constraint $\varepsilon_{k}=0$ defines exceptional points. If we set all hopping parameters to real numbers and  $w_{2}v_{1}=w_{1}v_{2}$, SEPs will locate at $k_{EP}=\pm\arccos(a_{r}/b_{r})$. 
In this situation, $|a_{r}/b_{r}|\leqslant1$ is required for the existence of SEPs. To make the computation and discussion easier, we set the SEP to $k_{EP}=0\quad(k_{EP}=-\pi)$. This necessitates $a_{r}/b_{r}=-1,\quad(a_{r}/b_{r}=1)$, which is independent of $s$.
The real part of the eigen-energies is more relevant since fermions fill the vacuum according to their real energies, while imaginary energies are the inverse of quasi-particle lifetimes. Given  $u^{2},v_{1,2},w_{1,2}\in\mathbb{R}$, we obtain the following expression for real energy:
\begin{widetext}
\begin{equation}
\text{Re}(\varepsilon_{k})=\left\{a_{r}+b_{r}\cos k+\left[(a_{r}+b_{r}\cos k)^{2}+s^{2}\sin^{2}k \right]^{1/2} \right\} ^{1/2}
\end{equation}
\end{widetext}

\begin{figure}
\includegraphics[width=8.5cm]{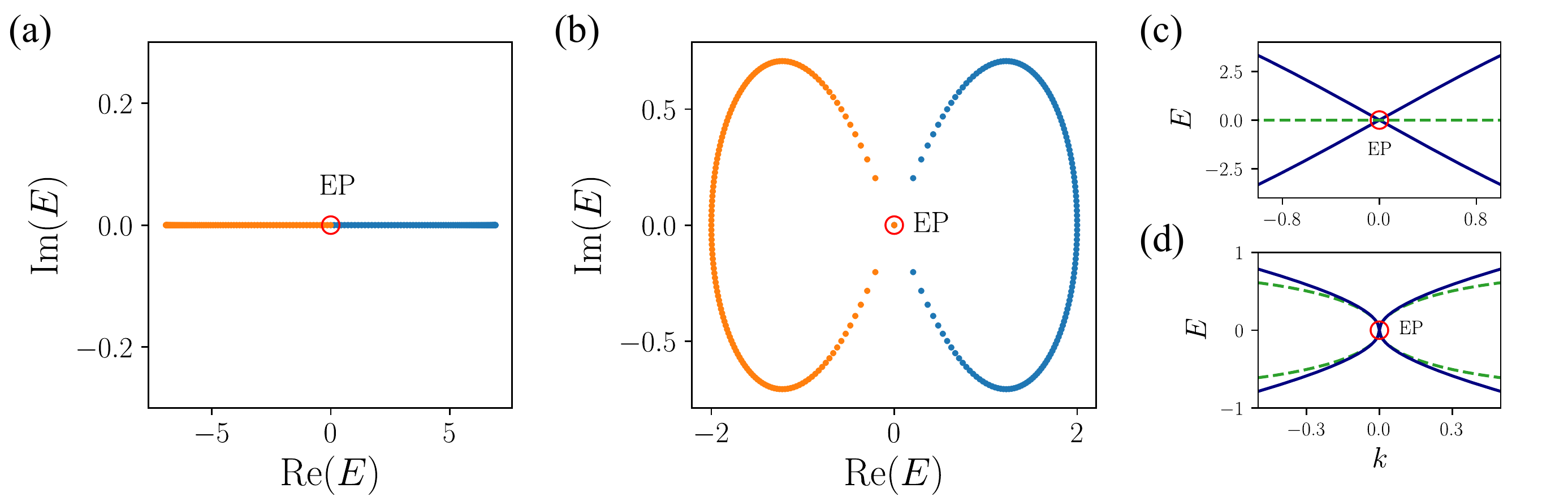}

\caption{\label{SPCT}The energy spectra of the model in PBC with fixed $a_{r}/b_{r}=1$:
(a) the cases with $s=0$; (b) the cases with $s\protect\neq0$ ($s=2$); (c)
the dispersion at the vicinity of EP for (a), solid line for real energy and dash line for imaginary energy; (d) the dispersion at the vicinity of EP for (b)}

\end{figure}

At the vicinity of SEPs, the real energy scales as $\text{Re}(\varepsilon_{k})\sim\{a_{r}k^{2}+[a_{r}^{2}k^{4}+s^{2}k^{2}]^{1/2}\}^{1/2}$,
it has a square-root dispersion, $\text{Re}(\varepsilon_{k})\sim\sqrt{k}$ for $s\neq0$ as shown in Fig.~\ref{SPCT}, which
is typical for 2-fold (second-order) exceptional points (EP2s). Specially,
it disperses as $\text{Re}(\varepsilon_{k})\sim k$ for $s=0$ and $a_{r}>0$ as shown in Fig.~\ref{SPCT}. 

The many-body ground state is constructed by filling all real negative modes. However, such filling is defective when the spectrum contains an EP because states at EPs are unphysical. To avoid this dilemma, an infinitesimal momentum shift is introduced to the band. Then the uppermost mode, i.e., the mode infinitely close to the EP, denoted as exceptional mode, displays exceptional properties through asymptotic behaviors.
As the $k$-power varies, the asymptotic behavior changes. The following section will examine the asymptotic behavior using the quantum entanglement measure.

\subsection{Type-I SEPs: $k$-linear dispersion\label{subsec:Linear}}

We take $-u^{2},v_{1,2},w_{1,2}>0$ and
\begin{equation}
    \label{lambda}
    w_{2}/w_{1}=v_{2}/v_{1}=\lambda
\end{equation}
i.e., $s=0$ for the first consideration.

We first analyze the symmetry of the model when $u=0$,  The Hamiltonian
is pseudo-Hermitian $U_{psH}H(\lambda)U_{psH}^{-1}=H^{\dagger}(\lambda)$,
with $U_{psH}=\textrm{Diag}\{1/\sqrt{\lambda},\sqrt{\lambda}\}$ ,
and (screw) $\mathcal{PT}$-symmetric $U_{\mathrm{PT}}H(\lambda)U_{\mathrm{PT}}^{-1}=H^{*}(\lambda),$
where the screw $\mathcal{PT}$-symmetry is unusually defined as
\begin{equation}
U_{\mathrm{PT}}=\left(\begin{array}{cc}
0 & 1/\sqrt{\lambda}\\
\sqrt{\lambda} & 0
\end{array}\right),U_{\mathrm{PT}}^{2}=1
\end{equation}
 which is non-symmorphic and returns $\sigma_{x}$ for $\lambda=1$.
The Hamiltonian has the Hermitian chiral symmetry $U_{Ch}H(\lambda)U_{Ch}^{-1}=-H(\lambda)$, 
where $U_{Ch}=\sigma_{z}$, which causes the eigen-energies appear in $\pm$ pairs, and time-reversal symmetry $\mathcal{T}H(k)\mathcal{T}^{-1}=H(-k)$,
with $\mathcal{T}$ is the normal time-reversal operator. 
Note that the Hamiltonian is similar to a Bloch Hamiltonian of the SSH model,
\begin{equation}
\label{sim}
S^{-1}H(\lambda)S=H_{\textrm{SSH}}(k),
\end{equation}
with similarity transformation,
\begin{equation}
    S=\left(\begin{array}{cc}
1 & 0\\
0 & \sqrt{\lambda}
\end{array}\right).
\end{equation}
This ensures that the Hamiltonian and the SSH model have the same spectrum.

While for imaginary $v_{1,2},w_{1,2}$, the Hamiltonian is anti-screw-$\mathcal{PT}$-symmetric
$U_{\mathrm{aPT}}H(\lambda)U_{\mathrm{aPT}}^{-1}=-H^{*}(\lambda),$
with anti-screw-$\mathcal{PT}$-symmetry is unusually defined as
\begin{equation}
U_{\mathrm{aPT}}=i\left(\begin{array}{cc}
0 & 1/\sqrt{\lambda}\\
\sqrt{\lambda} & 0
\end{array}\right),U_{\mathrm{aPT}}^{2}=-1
\end{equation}
It becomes anti-unitary $i\sigma_{x}$ when $\lambda=1$. 
The Hamiltonian still resembles an SSH model, albeit an imaginary one, with $H(\lambda)=S\widetilde{H}_{\textrm{SSH}}(k)S_{\mathrm{}}^{-1}.$ Then we have,
\begin{equation}
    U_{\mathrm{aPT}}S\widetilde{H}_{\textrm{SSH}}(k)S_{\mathrm{}}^{-1}U_{\mathrm{aPT}}^{-1}=-S\widetilde{H}_{\textrm{SSH}}^{*}(k)S_{\mathrm{}}^{-1},
\end{equation}
which reveals a "symmetry" transformation for $\widetilde{H}_{\textrm{SSH}}(k)$,
\begin{equation}
   \widetilde{U}_{\mathrm{aPT}}\widetilde{H}_{\textrm{SSH}}(k)\widetilde{U}_{\mathrm{aPT}}^{-1}=-\widetilde{H}_{\textrm{SSH}}^{*}(k), 
\end{equation}
with $\widetilde{U}_{\mathrm{aPT}}=S^{-1}U_{\mathrm{aPT}}S$ resembling a chiral symmetry for SSH model.

For $u\neq0$, the Hamiltonian has eigen-energies $\varepsilon_{k,\pm}=\pm[\lambda(w^{2}+v^{2})+u^{2}+2\lambda wv\cos k]^{1/2}$.
It remains pseudo-Hermitian, however, because all eigenvalues occur in real or complex conjugate pairs. The degeneracy continues. Additionally, the generalized  $\mathcal{PT}$-symmetry is preserved (as is the anti- $\mathcal{PT}$-symmetry for imaginary hoppings). Chiral symmetry, akin to particle-hole and time-reversal symmetry, becomes non-Hermitian (skew) chiral symmetry,
\begin{equation}
\label{nhch}
   \begin{array}{c}
U_{\textrm{nHCh}}H^{\dagger}(\lambda)U_{\textrm{nHCh}}^{-1}=-H(\lambda),\\
U_{\textrm{nHCh}}=\left(\begin{array}{cc}
1/\sqrt{\lambda} & 0\\
0 & -\sqrt{\lambda}
\end{array}\right)
\end{array}
\end{equation}
which still ensures that the eigen-energies emerge in pairs of $(\varepsilon_n,-\varepsilon_n) $.The unnormalized left and right eigenvectors are,
\begin{equation}
    \begin{array}{c}
\left|\tilde{\psi_{R}}\right\rangle _{k,\pm}=\left(\frac{\varepsilon_{k,\pm}+u}{we^{ik}+v},1\right)^{\textrm{T}},\\
\left|\tilde{\psi_{L}}\right\rangle _{k,\pm}=\left(\frac{\varepsilon_{k,\pm}-u}{\lambda(we^{ik}+v)},1\right)^{\textrm{T}}.
\end{array}
\end{equation}
To satisfy the bi-orthogonal condition, we should normalize the left
and right-left eigenvectors. We set $|\psi_{R,L}\rangle _{k,\pm}=|\tilde{\psi}_{R,L}\rangle _{k,\pm}/\sqrt{\langle \tilde{\psi}_{L}\mid\tilde{\psi}_{R}\rangle _{k,\pm}}$
as the normalized left and right eigenvectors.
The presence of SEPs necessitates that $\lambda(w-v)^{2}\leqslant-u^{2}\leq\lambda(w+v)^{2}$, at which point the norm $\langle \tilde{\psi}_{L}\mid\tilde{\psi}_{R}\rangle _{k,\pm}$ vanishes and the normalized left and right eigenvectors become ill-defined. Notably, when $u=0$, the SEP retrogrades to a diabolic point~\cite{yarkony} (level crossing point ) where algebraic singularities do not exist. Regarding entanglement entropy, it behaves similarly to a single-band free fermion chain in PBC. 

The entanglement entropy is unaffected by similarity transformations of the Bloch Hamiltonian, except for some boundary terms and finite size effects, as we shall see in the sections that follow. The similarity transformation can be considered to be an inserted metric in fermion fields, contributing only a metric-dependent factor to the partition function $Z_n$ and not to the central charge because the factor will be canceled out when the entropy is calculated as  $S^{(n)}=(1/(1-n))\text{ln}(Z_n/Z^n_1)$.

We first consider the specific case $u=\pm i\sqrt{\lambda}(w\pm v)$, which falls in the $\mathcal{PT}$-symmetry (anti-$\mathcal{PT}$-symmetry) 
breaking point of the model Eq.~\eqref{Hk}. This is also a case of pseudo-Hermiticity breaking, as its associated similarity transformation matrix into a Hermitian counterpart is losing its positivity. The SEP locates at $k_{EP}=-\pi$, around which
the wavefunctions are,
\begin{widetext}
\begin{equation}
\label{wavf1}
\begin{aligned} & |\psi_{R}\rangle _{k,\pm}=\frac{1}{\sqrt{2}}\left(\begin{array}{c}
i\textrm{sgn}(v-w)\textrm{ sgn}(-iu)\\
1
\end{array}\right)+\textrm{sgn}(-iu)\left(\begin{array}{c}
\pm\frac{\sqrt{vw}}{v-w}+\frac{w\textrm{sgn}(v-w)}{v-w}\\
0
\end{array}\right)\delta\kappa+O\left((\delta\kappa)^{2}\right)\\
 & |\psi_{L}\rangle _{k,\pm}=\frac{1}{\sqrt{2}}\left(\begin{array}{c}
-i\textrm{sgn}(v-w)\textrm{ sgn}(-iu)\\
1
\end{array}\right)+\textrm{sgn}(-iu)\left(\begin{array}{c}
\pm\frac{\sqrt{vw}}{v-w}-\frac{w\textrm{sgn}(v-w)}{v-w}\\
0
\end{array}\right)\delta\kappa+O\left((\delta\kappa)^{2}\right)
\end{aligned}
\end{equation}
\end{widetext}
The norm is $\langle \tilde{\psi}_{L}\mid\tilde{\psi}_{R}\rangle _{k,\pm}=\pm\frac{2ivw}{\sqrt{(v-w)^{2}}\sqrt{vw}}\delta\kappa+O((\delta\kappa)^{2}).$
As $k$ approaches the exceptional point, i.e., $\delta\kappa\rightarrow0$,
it is noticeable that the two eigenvectors coalesce into a single one
$(\pm i,1)^{\textrm{T}}$ with a positive group velocity
for $w>v$ while negative for $w<v$. For the second term in Eq.~\eqref{wavf1}, if we change the signs of $w,v$ from
the positive to negative, the entanglement features retain. 

 When $u=-i\sqrt{\lambda}(w-v)$ and $|v|>|w|$, which corresponds to a spontaneous $\mathcal{PT}$-symmetry breaking point, the spectrum is entirely real. There is a fixed EP at $k_{EP}=-\pi$, where the wavefunction is defective.
Around EPs, the energy is expressed as,
\begin{equation}
    \varepsilon_{k,\pm}=\pm2\sqrt{\lambda wv}\sin(k/2)\propto\pm\sqrt{\lambda wv}k.
\end{equation}

The correlation matrix (correlation functions in matrix form) is a block-Toeplitz matrix in the continuum limit, which ensures that the entanglement entropy has a vanishing linear scaling. We concentrate on a $2\times2$ block
\begin{equation}
\label{CMb}
\mathcal{C}_{\textrm{block}}=\left(\begin{array}{cc}
\mathcal{C}^{AA} & \mathcal{C}^{AB}\\
\mathcal{C}^{BA} & \mathcal{C}^{BB}
\end{array}\right),
\end{equation}
where $\mathcal{C}^{AA},\mathcal{C}^{AB},\mathcal{C}^{BA},\mathcal{C}^{BB}$
are correlators of forms $\langle c_{A(B)}^{\dagger}c_{A(B)}\rangle $ and A(B) is the sublattice lable.
We add a momentum cut-off $\delta_\kappa$ to the SEP to prevent divergence around SEPs, which
is equivalent to a tiny twist (a tiny flux insert) or tiny gap (the
gap size is $\Delta_{\delta\kappa}=\sqrt{\lambda wv}\delta\kappa$), then the matrix elements
divergent with $\delta\kappa$, $\mathcal{C}^{AA},\mathcal{C}^{AB},\mathcal{C}^{BA},\mathcal{C}^{BB}$
divergent as $1/\delta\kappa$. The signs of the determinant of the matrix (\ref{CMb})
in the limit of $\delta\kappa\rightarrow0$, denoted as $\textrm{Ind[\ensuremath{\mathcal{C}_{\textrm{block}}^{0}}]}$
can detect the entanglement feature as shown in Tab.~\ref{tab:cases}. We split the system in PBC into two continuous regions in real space to calculate the entanglement entropy.

\begin{figure}
\includegraphics[width=8.5cm]{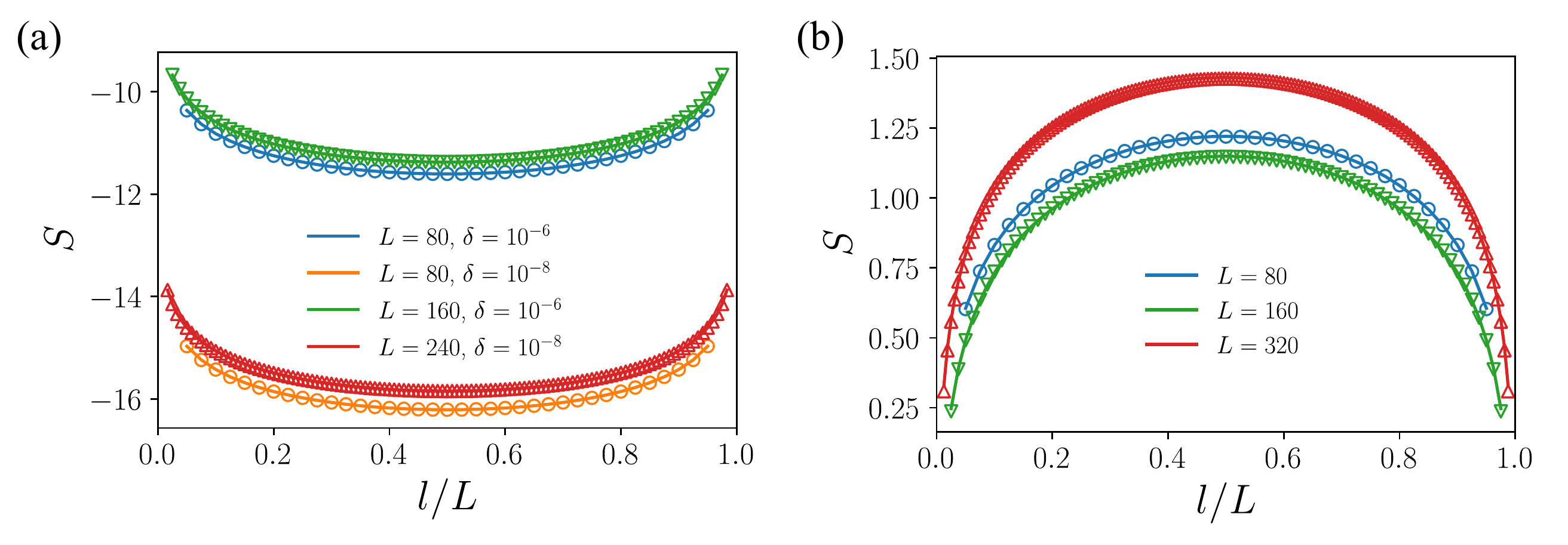}
\caption{\label{fig:Sc-2}(a) The entanglement entropy of choosing $\lambda=2,w=2,v=3,u=i$
in PBC, the fitting of Eq.~\eqref{eq:entropy} reveals
$c=-2.005,-2.005,-2.003,-2.002$ for blue, orange, green and red data. (b) The entanglement entropy of choosing $\lambda=2,w=2,v=3,u=i$ in APBC, the fitting of Eq.~\eqref{eq:entropy} reveals $c=0.995,1.064,1.021$ for blue, green and red data  \footnote{The non-vanishing small deviation of $c$ from exact $1$ at different $L$ is due to the presence of exceptional bound states, which is $L$-dependent. }. 
}
\end{figure}

\begin{figure}
\includegraphics[width=8.5cm]{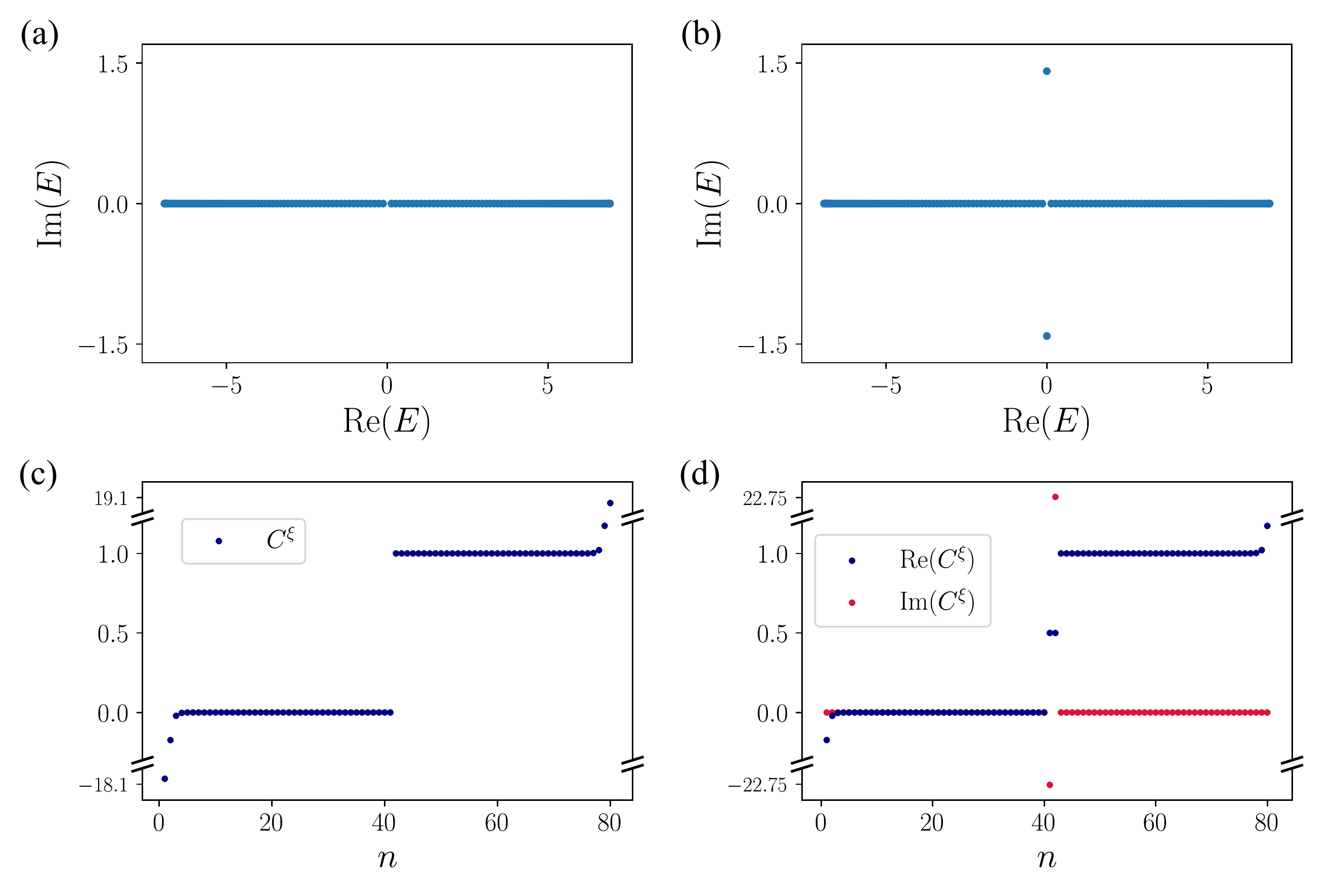}

\caption{\label{fig:Cspectrum a}Difference between non-universal cases and
$c=-2$ case. (a) The energy spectrum of $c=-2$ cases in OBC with
no boundary mode. (b) The energy spectrum of non-universal cases in OBC with two boundary states. (c) The eigenspectrum of the correlation
matrix in (a) with only real values. (d) The eigenspectrum of the correlation matrix in (a) with a pair of complex values.}
\end{figure}

In the case of $u=-i\sqrt{\lambda}(w-v)$ and $|v|>|w|$, nontrivial pairwise eigenvalues of the correlation matrix are
real but not falling in ordinary range $[0,1]$, with eigenvalue pairs $C^{\alpha}=1-C^{\beta}>1$ as shown in Fig.~\ref{fig:Cspectrum a}.
These pairs of eigenvalues contribute to negative entanglement entropy.
As shown in Fig.~\ref{fig:Sc-2}, the entanglement entropy scales logarithmically but with a negative coefficient in front of the logarithmic term
\begin{equation}
S(l)=\frac{c}{3}\ln\left(\frac{L}{\pi}\sin\frac{\pi l}{L}\right)+\textrm{const.}, \label{eq:entropy}
\end{equation}
where $c$ is the central charge of the corresponding conformal field theory
(CFT), the size of the subsystem is denoted by $l$, and the subleading term is a constant. $c$ is constant when $\delta\kappa$ is much lower than the unit moment $\frac{2\pi}{L}.$
The fact that the central charge is negative indicates that the CFT is non-unitary. The model could represent a bc-ghost (symplectic fermion) CFT with $c=-2$~\cite{kausch_curiosities_1995}. Note that the coefficient estimated from entanglement entropy is the effective central charge $c_{\textrm{eff}}$,
which relates to the true CFT central charge c as $c_{\textrm{eff}}=c-24\Delta_{\textrm{min}}$,
where $\Delta_{\textrm{min}}$ is the lowest conformal weight. 

To further verify the underlying symplectic fermionic structure of
the low-energy field theory, we try to extract the scaling dimension data of excitations. For unitary CFTs, the conformal vacuum, i.e.,
the ground state has the minimal conformal weight $L_{0}|GS\rangle=\Delta|GS\rangle$,
$\Delta=0$. And for all excited states, $\Delta>0$. While for a
non-unitary CFT, the physical vacuum with the lowest conformal weight may
not coincide with the $\Delta=0$ conformal vacuum. The ground state
$\Delta<0$. For a symplectic fermion CFT ($c=-2$) with zero modes i.e., in periodic boundary
condition (PBC), the lowest conformal weight $\Delta_{\textrm{min}}=0$,
thus $c_{\textrm{eff}}=c=-2$. The physical ground state respect conformal symmetry.
For anti-periodic boundary condition
(APBC), the lowest conformal weight $\Delta_{\textrm{min}}=-1/8$,
thus $c_{\textrm{eff}}=c-24\Delta_{\textrm{min}}=1$. 

To verify this,
we calculate the entanglement entropy under APBC, it is $S(l)=\frac{c_{\textrm{eff}}}{3}\ln\left(\frac{L}{\pi}\sin\frac{\pi l}{L}\right)+\textrm{const.}$,
with $c_{\textrm{eff}}=1$ as shown in Fig.~\ref{fig:Sc-2}. It is
noticeable that the model under APBC escapes the exceptional point and
behaves much like a Hermitian free (Dirac) fermion (CFT with $c=1$).

We can now investigate excitations whose conformal weights can be calculated by doubling the scaling dimension of chiral descendent states, with $\Delta_{\alpha,\textrm{PBC}}=2(E_{\alpha,\textrm{PBC}}-E_{0,\textrm{PBC}})/(E_{T}-E_{0,\textrm{PBC}})$, where $E_{\alpha,\textrm{PBC}}$, $E_{0,\textrm{PBC}}$ denote the chiral descendent state and ground state energy in PBC, respectively, and $E_{T}$ denotes the moment-energy tensor state energy. The moment-energy tensor state is equivalent to adding two minimal excitations above the zero-point energy, which always has a scaling dimension of $\Delta_{T}=2$.
In APBC, $\Delta_{\alpha,\textrm{APBC}}=2(E_{\alpha,\textrm{APBC}}-E_{0,\textrm{PBC}})/(E_{T}-E_{0,\textrm{PBC}})$.
For PBC the quasiparticle momentum take $k_{n}=-\pi+\dfrac{2\pi}{L}n$,
$n\leq L-1$, $n\in\mathbb{N}$. In APBC, the quasi-particle momentum
take $k_{n}=-\pi+\dfrac{2\pi}{L}(n+\dfrac{1}{2})$, $n\leq L-1$,
$n\in\mathbb{N}$. Therefore, the conformal tower for this model
in PBC is calculated as:
\begin{equation}
    \Delta_{n,\textrm{PBC}}=n,n\in\mathbb{N},
\end{equation}
and in APBC:
\begin{equation}
    \begin{array}{c}
\Delta_{0,\textrm{APBC}}=-\dfrac{1}{8},\Delta_{1,\textrm{APBC}}=\dfrac{3}{8},\\
\Delta_{n,\textrm{APBC}}=-\dfrac{1}{8}+\dfrac{1}{2}n,n\geq3,n\in\mathbb{N}.
\end{array}
\end{equation}

The symplectic fermion can also be interpreted using $\eta\xi$-ghost theory.
The $\eta\xi$-ghost CFT has two fermionic fields $\eta$ and $\xi$, with conformal
dimension respectively $\Delta_{\eta}=1$ and $\Delta_{\xi}=0$. 

The $\eta\xi$-ghost CFT action is 
\begin{equation}
    S=\frac{1}{2\pi}\int\mathrm{d}^{2}z(\eta\bar{\partial}\xi+\bar{\eta}\partial\bar{\xi}),
\end{equation}

with mode expansion: $\xi(z)=\sum_{n\in\mathbf{Z}}\xi_{n}z^{-n},\quad\eta(z)=\sum_{n\in\mathbf{Z}}\eta_{n}z^{-n-1}$
and fermionic anti-commutating relations $\{ \xi_{m},\eta_{n}\} =\delta_{m+n,0}$.
The operator's product expansion is 
\begin{equation}
    \begin{array}{c}
\xi(z)\eta(w)=\eta(z)\xi(w)=\frac{1}{z-w}+O(1),\\
\xi(z)\xi(w)=\eta(z)\eta(w)=O(1).
\end{array}
\end{equation}
 
And the energy-moment tensor
is
\begin{equation}
    T(z)=-:\eta(z)\partial\xi(z):=\sum_{n\in\mathbf{Z}}L_{n}z^{-n-2}.
\end{equation}

The Hamiltonian 
\begin{equation}
    H\propto L_{0}=-\sum_{n\in\mathbf{Z}}n:\eta_{n}\xi_{n}:
\end{equation}
is composed of two distinct classes of fermionic operators $\eta_{n},\xi_{n}$.
This is comparable to the left and right fermionic operators $\psi_{L,\alpha}^{\dagger}$
and $\psi_{R,\alpha}^{\dagger}$, respectively.

The low-energy theory of Hamiltonian Eq.~\eqref{hk} around spectral EP here can
be written as $H=\sum_{k}k\psi_{R,k}^{\dagger}\psi_{L,k}$, where we set
the Fermi velocity to 1. We calculate the corresponding correlators:
\begin{equation}
    \langle \psi_{L}^{\dagger}(x)\psi_{R}(0)\rangle =\frac{1}{L}\sum_{k}e^{ikx}=\frac{1}{\pi}\frac{\sin\pi x}{x}\propto\frac{1}{x},
\end{equation}

and the conjugate
\begin{equation}
    \langle \psi_{R}^{\dagger}(0)\psi_{L}(x)\rangle =\frac{1}{\pi}\frac{\sin\pi x}{x}\propto\frac{1}{x}.
\end{equation}
Also, note that
\begin{equation}
    \langle \psi_{L,R}^{\dagger}(x)\psi_{L,R}(0)\rangle \sim e^{ikx}C_{\textrm{div}}(\delta\kappa)\propto\dfrac{1}{x^{0}},
\end{equation}

where $C_{\textrm{div}}(\delta\kappa)$ 
is constantly divergent with
$\delta\kappa$. Thus, we obtain the scaling dimension 1 and 0 of ghost fermionic
operators respectively.

\begin{figure}
\includegraphics[width=8.5cm]{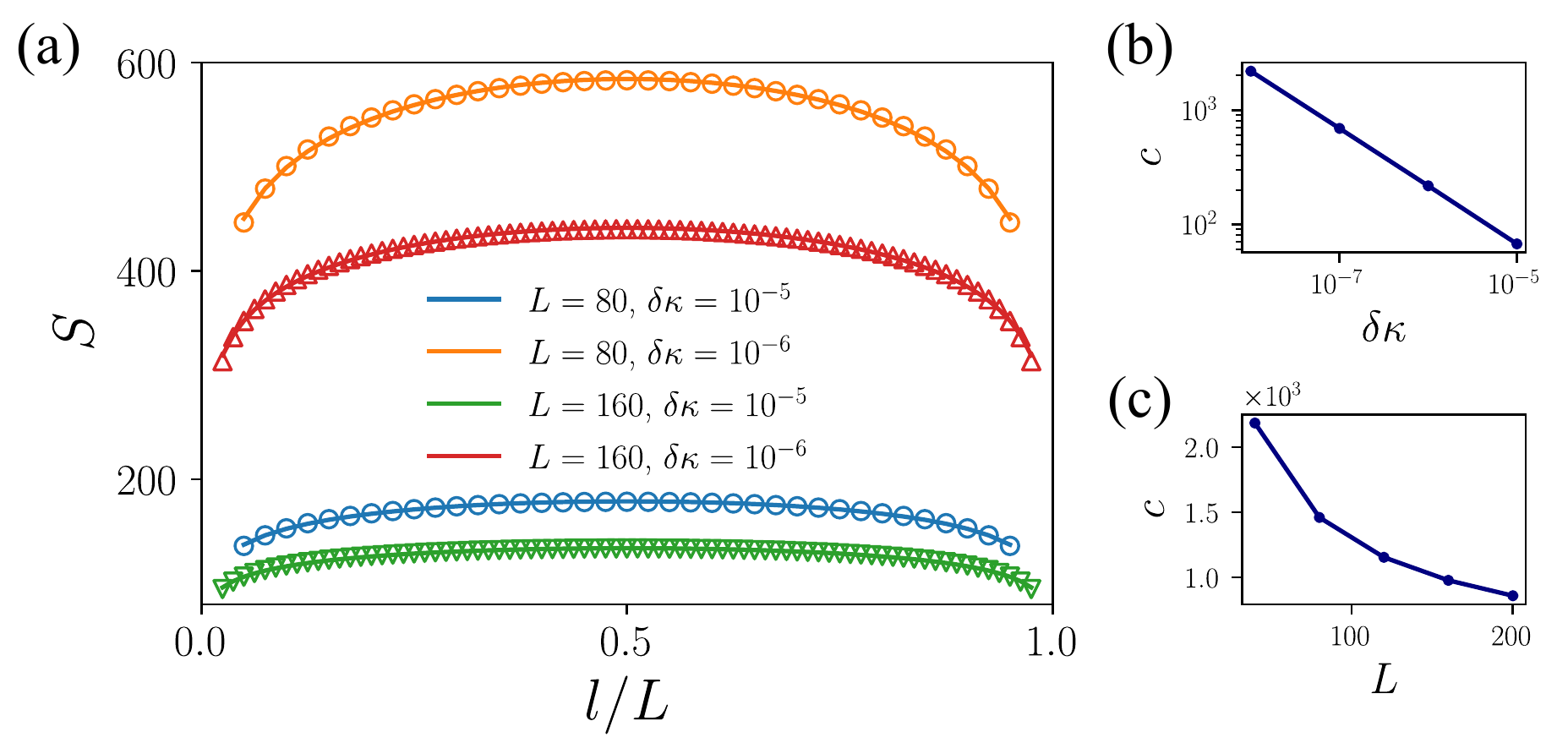}
\caption{\label{fig:Sa2}(a) The entanglement entropy of choosing $\lambda=1,w=3,v=2,u=i$ in PBC. The fitting of Eq.~\eqref{eq:entropy} reveals $c$ varies as $L$ and $\delta\kappa$, which is non-universal. (b) and (c) The $c$ dependence of $\delta\kappa$ and $L$.  }
\end{figure}

For $u=i\sqrt{\lambda}(w-v)$ and $|w|>|v|$
, it is a different story. Although it corresponds to a $\mathcal{PT}$-symmetry spontaneous breaking point, it has additional topological boundary modes. There are also an extra pair of nontrivial eigenvalues $C^{\textrm{ntr,}\pm}=0.5\pm i\varUpsilon$,
where the imaginary part $\varUpsilon$ is found to depend on not only
subsystem length $l$ but also the moment distance $\delta\kappa$
and the total system size $L$. 

These eigenvalues ensure that entanglement entropy is positive.
$S=\gamma\ln[\sin(\pi l/L)]+\textrm{const., }$ $\gamma>0$. However, as illustrated in Fig.~\ref{fig:Sa2}, the coefficient is not universal. The spectrum of this model in OBC is shown in Fig.~\ref{fig:Cspectrum a}(b). When $|w|>|v|$, $u=\pm i\sqrt{\lambda}(w-v)$, and $wv>0$, a pair of topological edge modes with imaginary energy $\pm u$ originates from the model's topology, which is protected by the non-Hermitian chiral symmetry defined in Eq.~\eqref{nhch}.
They are not exponentially localized at entanglement spectrum boundaries, despite the fact that they are physically localized states~\cite{chang_entanglement_2020}.
As a result, they provide an additional size-dependent contribution to entanglement entropy. There are also similar phenomena in gapless topological systems and symmetry-enhanced quantum critical systems, for which one can refer to~\cite{gapless17,gapless21}. In the thermodynamic limit, this additional pair of topological modes destroys the conformal (scale-invariant) symmetry (the fitting central charge goes to 0 as $L$ approaches infinity). Additionally, because exceptional modes are dependent on $\delta\kappa$, the hybridization of exceptional modes and non-Hermitian boundary modes results in a non-universal dependence on $\delta\kappa$ for the effective central charge. The entanglement spectrum is addressed in detail in Sec.~\ref{ESQ}.

If we consider that $w,v$ is purely imaginary, as shown in the Tab.~\ref{tab:cases}, the Hamiltonian has anti-$\mathcal{PT}$ symmetry, and the anti-$\mathcal{PT}$ symmetry breaking point contains spectral singularities. The dispersion is,
\begin{equation}
    \varepsilon_{k,\pm}=\pm2\sqrt{\lambda wv}\sin(k/2)\propto\pm\sqrt{\lambda wv}k.
\end{equation}

True spectra require a value of $wv>0$, i.e., $\textrm{Im}(w)\textrm{Im}(v)<0$.
Tab.~\ref{tab:cases} displays the detailed results.\footnote{Note that for imaginary spectra with anti-$\mathcal{PT}$-symmetry.
The imaginary spectra play the role as the real spectra in the above cases, which can be regarded as an imaginary mirror of $\mathcal{PT}$-symmetric
cases.}

\begin{table*}
\begin{centering}
\begin{tabular}{c|c|c|c|c|c|c|c}
\hline 
Parameters & \multicolumn{1}{c|}{Choice of $u$} & Sign of $wv$ & Spectrum & $w$ and $v$ & Entanglement Entropy & $\textrm{Ind[\ensuremath{\mathcal{C}_{\textrm{block}}^{0}}]}$ & Edge Modes\tabularnewline
\hline 
\hline 
\multirow{6}{*}{$\begin{array}{c}
w,v\textrm{ real};\\
u\textrm{ imaginary}
\end{array}$} & \multirow{3}{*}{$\begin{array}{c}
u=\pm i\sqrt{\lambda}(w-v);\\
k_{\textrm{EP}}=-\pi
\end{array}$} & \multirow{2}{*}{$wv>0$} & \multirow{2}{*}{real spectra} & \multicolumn{1}{c|}{$\left|w\right|<\left|v\right|$} & $c=-2$ & +1 & None\tabularnewline
\cline{5-8} \cline{6-8} \cline{7-8} \cline{8-8} 
 &  &  &  & $\left|w\right|>\left|v\right|$ & non-universal & -1 & $\pm u$\tabularnewline
\cline{3-8} \cline{4-8} \cline{5-8} \cline{6-8} \cline{7-8} \cline{8-8} 
 &  & $wv<0$ & imaginary spectra & all & $c=-2$ & +1 & None/Indiscernible\footnote{when $\left|w\right|>\left|v\right|$, there are still topological boundary modes with energies $\pm u$ in OBC. On the complete imaginary or real spectrum, however, the boundary modes become indistinguishable, and they are also unprevalent on the entanglement spectrum. In the main text, we regard the indiscernible modes to be non-existent.}\tabularnewline
\cline{2-8} \cline{3-8} \cline{4-8} \cline{5-8} \cline{6-8} \cline{7-8} \cline{8-8} 
 & \multirow{3}{*}{$\begin{array}{c}
u=\pm i\sqrt{\lambda}(w+v);\\
k_{\textrm{EP}}=0
\end{array}$} & $wv>0$ & imaginary spectra & all & $c=-2$ & +1 & None/Indiscernible\tabularnewline
\cline{3-8} \cline{4-8} \cline{5-8} \cline{6-8} \cline{7-8} \cline{8-8} 
 &  & \multirow{2}{*}{$wv<0$} & \multirow{2}{*}{real spectra} & $\left|w\right|<\left|v\right|$ & $c=-2$ & +1 & None\tabularnewline
\cline{5-8} \cline{6-8} \cline{7-8} \cline{8-8} 
 &  &  &  & $\left|w\right|>\left|v\right|$ & non-universal & -1 & $\pm u$\tabularnewline
\hline 
\multirow{6}{*}{$\begin{array}{c}
w,v \textrm{ imaginary};\\
u\textrm{ real}
\end{array}$} & \multirow{3}{*}{$\begin{array}{c}
u=\pm i\sqrt{\lambda}(w-v);\\
k_{\textrm{EP}}=-\pi
\end{array}$} & $wv>0$ & real spectra & all & $c=-2$ & +1 & None/Indiscernible\tabularnewline
\cline{3-8} \cline{4-8} \cline{5-8} \cline{6-8} \cline{7-8} \cline{8-8} 
 &  & \multirow{2}{*}{$wv<0$} & \multirow{2}{*}{imaginary spectra} & $\left|w\right|<\left|v\right|$ & $c=-2$ & +1 & None\tabularnewline
\cline{5-8} \cline{6-8} \cline{7-8} \cline{8-8} 
 &  &  &  & $\left|w\right|>\left|v\right|$ & non-universal & -1 & $\pm u$\tabularnewline
\cline{2-8} \cline{3-8} \cline{4-8} \cline{5-8} \cline{6-8} \cline{7-8} \cline{8-8} 
 & \multirow{3}{*}{$\begin{array}{c}
u=\pm i\sqrt{\lambda}(w+v);\\
k_{\textrm{EP}}=0
\end{array}$} & \multirow{2}{*}{$wv>0$} & \multirow{2}{*}{imaginary spectra} & $\left|w\right|<\left|v\right|$ & $c=-2$ & +1 & None\tabularnewline
\cline{5-8} \cline{6-8} \cline{7-8} \cline{8-8} 
 &  &  &  & $\left|w\right|>\left|v\right|$ & non-universal & -1 & $\pm u$\tabularnewline
\cline{3-8} \cline{4-8} \cline{5-8} \cline{6-8} \cline{7-8} \cline{8-8} 
 &  & $wv<0$ & real spectra & all & $c=-2$ & +1 & None/Indiscernible\tabularnewline
\hline 
\end{tabular}
\par\end{centering}
\caption{\label{tab:cases}Various choices of parameters $w,v,u$ and their
corresponding properties.}

\end{table*}

When $|u|>\sqrt{\lambda}|w-v|$, the energy spectrum is neither completely real nor completely imaginary. When $k<k_{e-}=-\arccos\text{\{[}\lambda(w^{2}+v^{2})+u^{2}]/2\lambda wv\}$
or $k>k_{e+}=\arccos\text{\{[}\lambda(w^{2}+v^{2})+u^{2}]/2\lambda wv\}$, the energy spectrum is fully imaginary, occurring in PT-broken phase. The dispersion around SEPs is $\varepsilon_{k,\pm}\propto\pm\sqrt{k}$ for $k_{EP-}<k<k_{EP+}$ and $\varepsilon_{k,\pm}\propto\pm i\sqrt{k}$ otherwise. Since discrete modes are not certain to be located on the SEPs, if we continue to construct the ground state by half-filling the lowest states with real energies, i.e., by including some purely imaginary states, the correlation matrix will yield complex eigenvalues and thus contribute to complex entanglement entropy without exhibiting logarithm scaling behavior.
When modes do not locate at SEPs, the dispersion around the mode is $\varepsilon_{k,\pm}\propto\pm k$. If we only include modes with real negative energies in the Dirac sea, the entanglement entropy will be similar to that of the free (Dirac) fermion CFT.
When an imaginary mode is added to the ground state, the correlation matrix yields complex eigenvalues and the entanglement entropy becomes complex, which means any modes with imaginary energies will immediately destroy the vacuum's conformal symmetry and drive relevantly the system to other exotic "phases". When certain parameters are set in such a way that discrete modes precisely locate at SEPs, the exact conformal symmetry is also broken. Due to the inclusion of the mode surrounding one SEP in the ground state, the entanglement entropy is complex and scales logarithmically, which falls under the category of the situation discussed in the section below. 


It is notable that if we fill the ground state far from SEPs, the entanglement entropy gives a free Dirac fermion CFT behavior. This indicates the essence of exceptional points-related states on the entanglement features of the constructed ground states.

\subsection{Type-II SEPs: $k$-square-root dispersion \label{srd}}

When $w_{2}v_{1}\neq w_{1}v_{2}$, i.e., $s\neq0$, $v_{1,2},w_{1,2}\in\mathbb{R}$,
the Hamiltonian is no longer quasi-Hermitian, resulting in a complex spectrum and a non-linear dispersion around EPs.
 For instance, if
we take $w_{1}w_{2}+v_{1}v_{2}+u^{2}=w_{2}v_{1}+w_{1}v_{2}>0$,  the EP locates at $k_{EP}=-\pi$, around which dispersion is complex,
\begin{equation}
    \varepsilon_{k}\approx\sqrt{i(w_{1}v_{2}-w_{2}v_{1})}\sqrt{k}.
\end{equation} 

The wavefunctions are
\begin{widetext}
\begin{equation}
\begin{aligned} & |\psi_{R}\rangle _{k,\pm}=\left(\begin{array}{c}
\frac{\sqrt{-(v_{1}-w_{1})(v_{2}-w_{2})}}{v_{2}-w_{2}}\\
1
\end{array}\right)+\left(\begin{array}{c}
\pm\frac{\sqrt{i(w_{1}v_{2}-w_{2}v_{1})}}{v_{2}-w_{2}}\\
0
\end{array}\right)\sqrt{\delta\kappa}+O\left(\delta k\right)\\
 & |\psi_{L}\rangle _{k,\pm}=\left(\begin{array}{c}
\frac{\sqrt{-(v_{1}-w_{1})(v_{2}-w_{2})}}{v_{1}-w_{1}}\\
1
\end{array}\right)+\left(\begin{array}{c}
\pm\frac{\sqrt{-i(w_{1}v_{2}-w_{2}v_{1})}}{v_{1}-w_{1}}\\
0
\end{array}\right)\sqrt{\delta\kappa}+O\left(\delta k\right)
\end{aligned}
\end{equation}
\end{widetext}

\begin{figure}
\includegraphics[width=8.5cm]{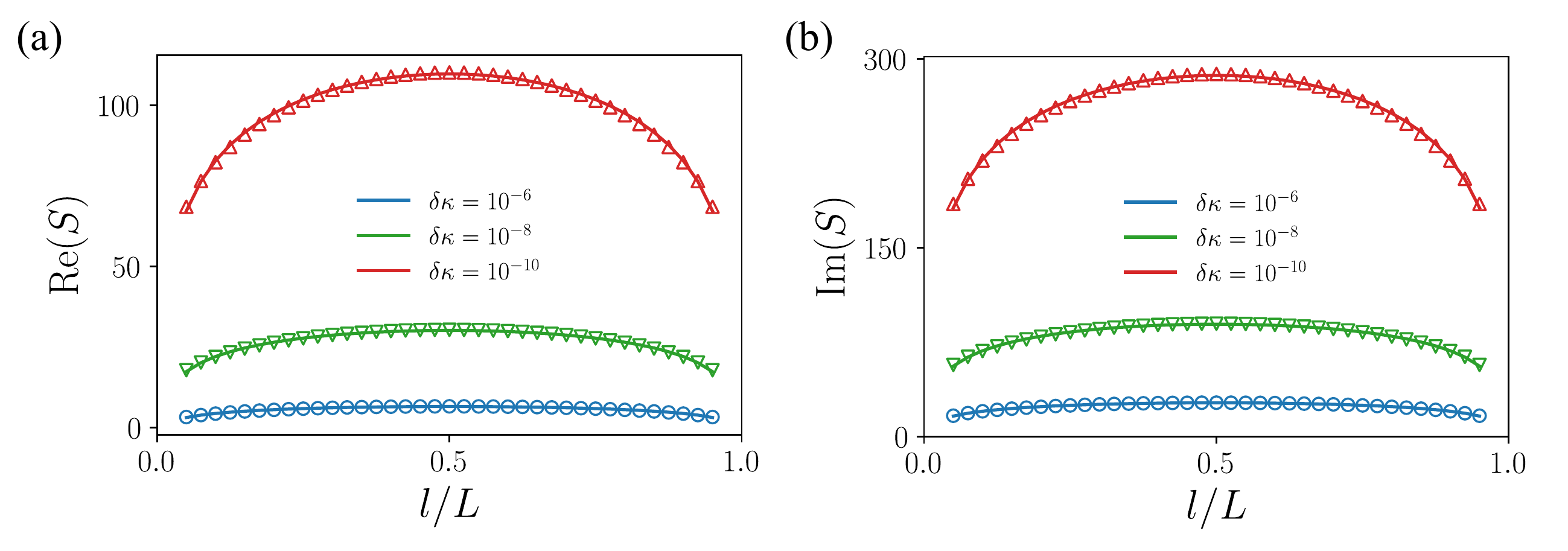}

\caption{\label{fig:sb1}(a) The entanglement entropy scaling of choosing $w_1=1, v_1=1, w_2=0.3, v_2=0.7$ in PBC. The fitting of Eq.~\eqref{eq:entropy} reveals
$c=5.5+17.0 i$ (the blue), $c=20.7+53.9 i$ (the green) and $c=68.8+170.1 i$ (the red) with total size $L=80$.  }
\end{figure}

The norm $\langle \tilde{\psi}_{L}\mid\tilde{\psi}_{R}\rangle _{k,\pm}=\mp\frac{\sqrt{i(w_{1}v_{2}-w_{2}v_{1})}}{\sqrt{-(v_{1}-w_{1})(v_{2}-w_{2})}}\sqrt{\delta\kappa}+O\left(\delta k\right)$ verges on zero as $\sqrt{\delta\kappa}$.
In the scenario, The filled Fermi sea shows no conformal symmetry at first blush. However, when inserting a momentum shift $\delta\kappa$ or a complex gap 
\begin{equation}
\label{gap}
    \Delta_{\kappa}=\sqrt{i(w_{1}v_{2}-w_{2}v_{1})}\sqrt{\delta\kappa},
\end{equation} 
the dispersion around the shifted Fermi point is approximately linear $\textrm{Re}(\epsilon_{k}),\textrm{Im}(\ensuremath{\epsilon_{k}})\propto(1/2\sqrt{\delta\kappa})k$, preserving possible proximate conformal symmetry similar to the free fermion. 
The correlation matrix $\mathcal{C}^{AA}=\mathcal{C}^{BB}=1/2$ is independent of $\delta\kappa$, indicating that each site has an equal probability of occupation, while $\mathcal{C}^{AB}$ diverges as $1/\sqrt{\delta\kappa}$ and $\mathcal{C}^{BA}$ approaches 0 as $\sqrt{\delta\kappa}$. $\ensuremath{|\mathcal{C}_{\textrm{block}}^{0}|}=0$ does not convey information about the entanglement pattern anymore since it always equals to zero. The correlation matrix's nontrivial eigenvalues also occur in pairs with  $C^{\alpha}=1-C^{\beta}$, but they are both complex and dependent on $L$ and $\delta\kappa$. The entanglement entropy has the logarithm form $S=\gamma_{c}(\delta\kappa,L)\ln\left[\sin\left(\pi l/L\right)\right]+\textrm{const.}$, $\gamma_{c}(\delta\kappa,L)\in\mathbb{C}$, as illustrated in Fig.~\ref{fig:sb1}.
Since the entanglement property is not dependent on the chemical potential, we can take $u=0$. If $\lambda_{1}=w_{1}/v_{1}$
and $\lambda_{2}=w_{2}/v_{2}$ remain constant, the correlation functions stay unchanged.
To further minimize the number of parameters, we can set the SEP to $k_{\textrm{EP}}=-\pi$, which allows us to tweak the model's correlation and entanglement properties with a single parameter. 

In the case where $u=0$. $w_{1}=v_{1}$ and $w_{2}>v_{2}$, as $w_1$ varies and other parameters remain constant, the SEP ceases to exist and a real or imaginary gap opens. It resembles a critical point, which divides the system into phases with different topologies (different vorticities surrounding 2 EPs in k-space)~\cite{yin18}. However, under such parameter choices, the critical point does not involve any symmetry breaking or conventional topological quantum phase transitions~\cite{arouca_unconventional_2020}. It may mimic a first-order phase transition, which is a common phenomenon in $\mathcal{PT}$-broken systems~\cite{fruchart_non-reciprocal_2021}, where SEPs mark the first-order phase transition.

The drift complex logarithmic scaling entanglement entropy is reminiscent of complex conformal field theories (cCFTs) with complex central charges, which only have approximate conformal symmetry and are suggested to be connected to first-order weak phase transitions. Previous research has revealed the phenomena in various  classical statistical models and strongly correlated or disordered field theories~\cite{benini_conformality_2020,gorbenko_walking_2018,kaplan_conformality_2009,ma_shadow_2019}. In free fermion models, the phenomenon does not appear to be achievable.

However, the seemingly non-interacting non-Hermitian Hamiltonian can be written as $H_{\textrm{eff}}(k)=H_{0}(k)+\Sigma(\omega,k)$, where $H_{0}(k)$ is the true non-interacting Hermitian part and $\Sigma(\omega,k)$ is the self-energy from certain types of interactions or disorders, the imaginary part of which represents the lifetime of the quasi-particle. This makes the non-Hermitian Hamiltonian an effective model for a strongly interacting or disordered model in the sense of single-particle Green's function~\cite{kozii_non-hermitian_2017,michishita_equivalence_2020}.

\begin{figure*}
\includegraphics[scale=0.5]{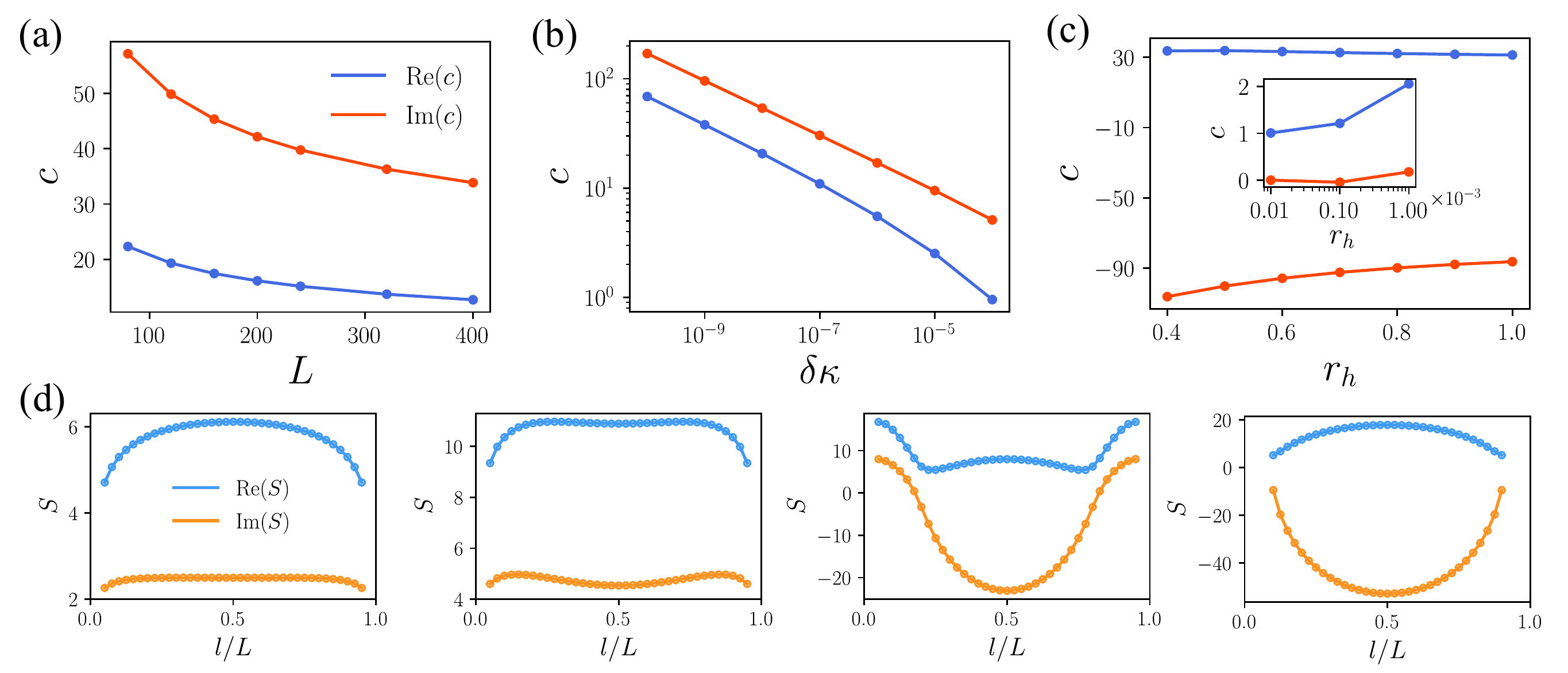}

\caption{\label{csvsp}(a) The fitting central charge versus $L$ with $\delta\kappa=10^{-8}$ fixed and $r_h=10^{6}$ ($+\infty$ fixed point). (b)  The fitting central charge versus $\delta\kappa$ with $L=80$ fixed and $r_h=10^{6}$ ($+\infty$ fixed point). (c)  The fitting central charge versus $r_h$ with $L=80$ and $\delta\kappa=10^{-8}$.  In the ultra-violet region, where the system is small in size, the central charge is highly dependent on the system size $L$; inner panel: $c$ versus $r_h$ when $r_h$ is much close to 0 (from $10^{-5}$ to $10^{-3}$). It is notable the $c$ is close to $1$ when $r_h=10^{-5}$. (d) The crossover of entanglement entropy from a complex one to a free fermion one under $L=80$ and $\delta\kappa=10^{-8}$, from left to right $r_h=0.01,0.1,0.2,0.4$. }
\end{figure*}

\begin{figure*}
\centering
\includegraphics[width=15cm]{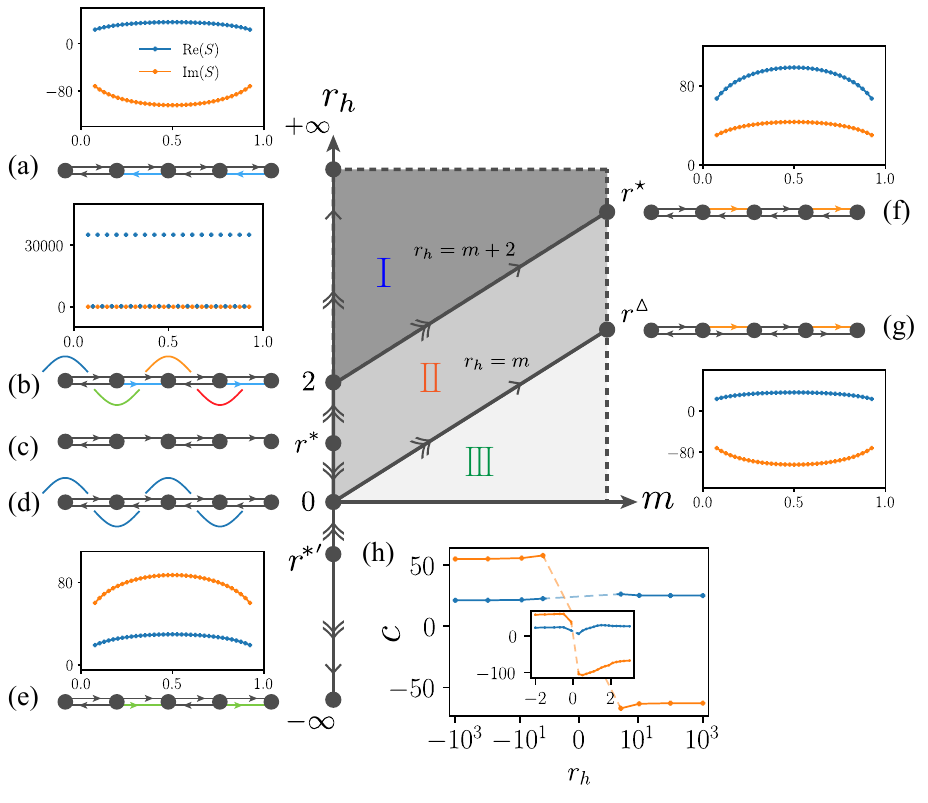}

\caption{\label{core} As shown in the central region of the figure, the "phase diagram" of model \eqref{fieldt} includes three phases: phase I is the real gapped phase (complex spectrum), while phases II and III are imaginary gapped phases with different $k$-locations of gap opening. The three exceptional lines $m=0$, $m=r_h$, and $m+2=r_h$ are marked with bold black lines, with each point on the line being an exceptional point. The possible flow on the line is indicated by arrows, where double arrows represent strong flow and single arrows represent weak flow. Possible fixed points are denoted by bold black dots, and their corresponding chain configurations and entanglement entropies (calculated with $L=80$ and $\delta\kappa=10^{-8}$) are labeled nearby. (a) shows the chain configuration at $r_h=+\infty$, where the light blue dashed line indicates a hopping bond that is much larger than others, and the corresponding entanglement entropy is located above it (calculated with $r_h=10^{5}$). (b) displays the chain configuration at $r_h=2$ (tricritical point), where the blue reverse arrow indicates a negative hopping coefficient with the same amplitude as others. The four different colored wavy lines indicate that the wave function actually contains a factor of $i^{n}$, and the corresponding entanglement entropy results are located above them. The entanglement entropy at $r_h=2$ is actually divergent, so we can only approach it infinitely closely (we take $r_h=2-10^{-16}$ for calculation), and a clear 4-periodic pattern can be observed. (c) shows the chain configuration at some intermediate fixed point $r^{*}$, where the central charge takes the maximum value. (e) At the point where $r_h=-\infty$, the chain configuration is depicted using a dashed green bond which indicates that the bond strength is much larger than other black hopping bonds, but the corresponding hopping coefficient is negative, hence the arrow direction is reversed. The entanglement entropy corresponding to this configuration is labeled above it (calculated with $r_h=-10^{5}$).
The chain configurations of two additional fixed points $r^{\star}$ and $r^{\vartriangle}$ at infinity are shown in (f) and (g), respectively. They are located at $m+2=r_h$ and $m=r_h$, respectively. The difference between them lies in the signs of the hopping parameters (indicated by the downward arrows) that commute with the infinite jump operator. Specifically, the downward arrows in (g) have opposite signs to those in (f). The entanglement entropies for these fixed points are labeled above and below the corresponding configurations.
It can be seen that the 1D chain configurations of the fixed points at $r^{*}$ and other infinite distances bear some similarity to the 1D deconfined quantum critical points (DQCP) in~\cite{huse_commensurate_1984,senthil_deconfined_2004,jiang_ising_2019,wang_deconfined_2017,roberts_deconfined_2019,huang_emergent_2019,liu_emergence_2022,zhang_exactly_2023}, and we believe that they still have approximate conformal symmetry. (h) displays the flow of the central charge on the $r_h$ axis, and the calculation result shows that the change of the central charge is no more than $10^{-8}$ when $|r_h|>10^3$. The central panel shows the flow of $c$ when $-2<r_h<2$, and the behavior near 0 can be seen in Figure \ref{csvsp}(c). Around $r_h=1.2$ and $r_h=-0.6$, $|c|$ takes a maximum value, corresponding to a repulsive fixed point. Of course, the calculated points depend on the choice of $L$ and $\delta\kappa$.}
\end{figure*}

In order to investigate the drifting complex entanglement entropy, we propose the following field-theoretic model:

\begin{widetext}
    \begin{equation}
    \label{fieldt}
H_{\textrm{nH}}\sim\int dx\left\{ i\bar{\Psi}(x)\gamma^{x}\partial_{x}\Psi(x)-m\bar{\Psi}(x)i\gamma^{5}\Psi(x)-ir_{h}\left[1+(-1)^{x}\right]\left[\psi^{\dagger}(x)\psi(x)-\psi^{\dagger}(x)\partial_{x}\psi(x)\right]\right\} ,
\end{equation}
\end{widetext}

where $\bar{\Psi}(x)$ and $\Psi(x)$ denote the Dirac fermions with mass $m$, while the last term is non-Hermitian and consists of chiral fermions represented by $\psi^{\dagger}(x)$ and $\psi(x)$. The non-Hermitian lattice realization of the chiral fermions can be found in~\cite{chen2023fate}. Its discrete form is equivalent to the non-Hermitian lattice model \eqref{tmodel} by setting $w_{2}=w_{1}=1$, $v_{1}=1+m$ and $v_{2}=1+m-r_{h}$. Hence, the phase diagram can be obtained as shown in Fig.~\ref{core} based on the spectrum properties.

The first term in the model \eqref{fieldt} represents a massless Dirac fermion field. The second term represents a mass term with lattice polarization, which is the leading relevant term that introduces an energy gap in the system. The last term, which is the chiral perturbation term, explicitly breaks the $\mathcal{PT}$-symmetry (either $\mathcal{P}$ or $\mathcal{T}$).  It is believed that the chiral perturbation term will drive the system towards a non-critical, gapless phase, as reported in several models. However,  it is challenging to address the chiral perturbation term directly in the non-Hermitian context using field-theoretic methods. Instead, we can estimate the renormalization group (RG) flow of the coupling coefficient $r_{h}$ by performing numerical calculations of the entanglement entropy through the c-theorem of conformal field theory (CFT).

To capture the key physics,  we focus our attention primarily on the positive $r_h$ axis and stable fixed points. We first examine the exceptional line with $m=0$. Around $r_{h}=0$, as shown in Fig.~\ref{csvsp}(d) the entanglement entropy displays a crossover behavior from a complex scaling to a $c=1$ scaling, indicating a marginal departure from the massless Dirac fermion. At $r_{h}=2$, the entanglement entropy shows tetramerization. This phenomenon can be demonstrated by analyzing the wavefunction. Specifically, the model at $r_{h}=2$ cannot be transformed to a Hermitian chain through the similarity transformation in open boundary conditions (OBC). However, it is noted that the wavefunction takes the form $\sim\beta_{(1,2)}^{n}$ with $\beta_{(1,2)}=i$ when $L/2+1$ is odd and $\beta_{(1,2)}=1,-1$ when $L/2+1$ is even. Therefore, as shown in Fig.~\ref{core}(b), a natural 4-period of wavefunction form emerges when we bipartite the system to compute the entanglement entropy.

\begin{figure}
\centering
\includegraphics[width=8.5cm]{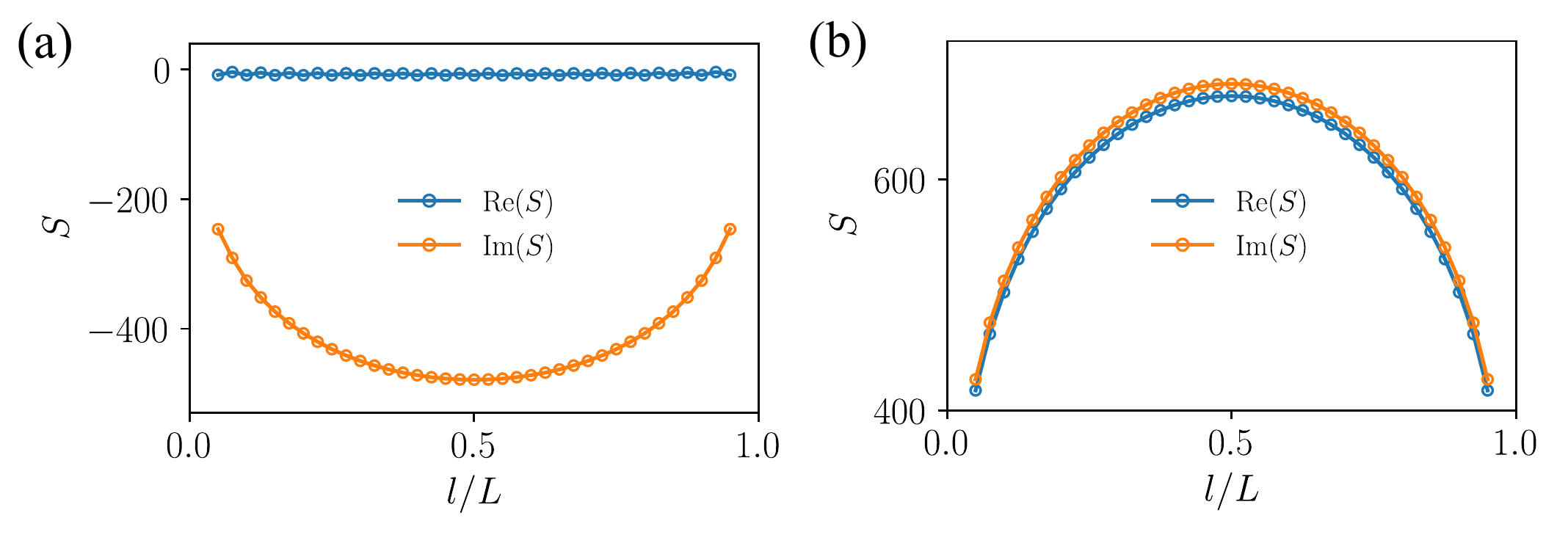}
\caption{\label{imd} the scaling behavior of entanglement entropy at $r_h\rightarrow\infty$, when the small gap is artificially set to be (a) pure imaginary or (b) pure real. The gap size is around $10^{-4}$, the lattice size is 80, and the momentum offset $\delta\kappa$ is fixed at $10^{-8}$.}
\end{figure}

As $r_{h}$ is increased to a large value, the flow of the central charge is extremely slow. According to c-theorem of conformal field theory (CFT), which states that:

\begin{equation}
\frac{\partial}{\partial r_h}c(\lambda)=-\frac{dr_h}{d\ln L}\label{eq:cthem-1}
\end{equation}

where $c(\lambda)$ is the central charge and $L$ is the system size, at infinite strong coupling $r_h\rightarrow\infty$, there exists a fixed point where the central charge reaches a particular value $c=c^*$. However, the strong scale dependence of the central charge can still be observed even at very large $r_h$. It is noteworthy that the momentum shift serves a dual purpose in theory. Primarily, it functions as an infrared cut-off, analogous to the system size $L$. Secondarily, it is incorporated into the theory as a mass term, as expressed in Eq.~\eqref{gap}. As demonstrated in Fig.~\ref{imd}, we artificially manipulate the gap induced by $\delta\kappa$ to be purely real or imaginary, resulting in distinct features in the entanglement entropy, thereby highlighting the role of $\delta\kappa$ as an additional complex mass term (complex artificial gap) in theory.

\begin{figure}
\centering
\includegraphics[width=8.5cm]{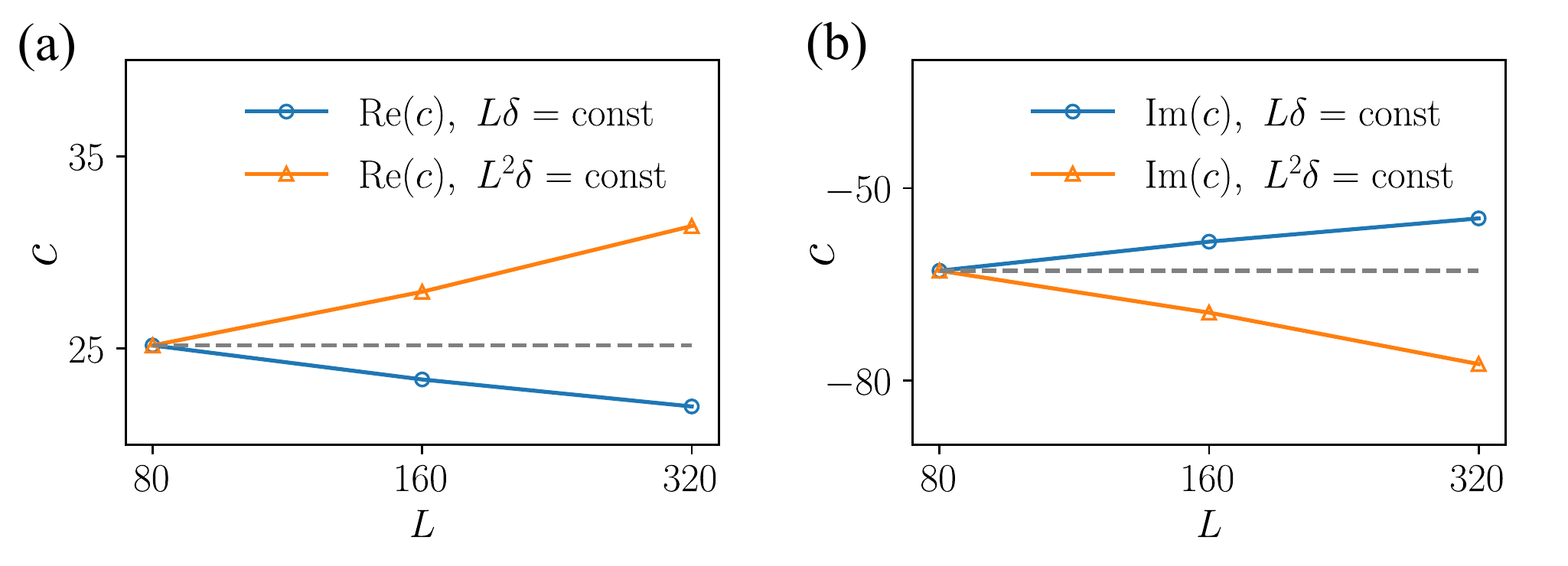}

\caption{\label{add}   the drift of the (a) real part and (b) imaginary part of the central charge as the system size increases, while keeping $L\delta\kappa$-invariant and $L^2\delta\kappa$-invariant at the fixed point $r_h\rightarrow\infty$. The dashed line corresponds to the central charge that does not drift with size and is expected to be $L^{\zeta}\delta\kappa$-invariant, where $1<\zeta<2$.}
\end{figure}

Due to the IR-ill-defined nature of the model both theoretically and numerically, the underlying physics of the fixed point can only be accessed asymptotically by letting $r_h$, $L$, and $\delta\kappa$ approach infinity and $\Delta\sim\delta\kappa r_h\rightarrow0$. However, the fixed point physics depends on the relationship between $L$ and $\delta\kappa$. Fig.~\ref{csvsp}(a),(b) and Fig.~\ref{add} show that $c(L,\delta\kappa)=c^{}(\infty,0)$ depends on the limiting sequence. For $L\gg1/\delta\kappa\rightarrow\infty$, $c^{}(\infty,\delta\kappa)\rightarrow0$, corresponding to a non-negligible gap, i.e., conformal symmetry breaking. For $L\ll1/\delta\kappa\rightarrow\infty,c^{}(L,0)\rightarrow\infty$, which retains conformal symmetry. The relationship between $\textrm{Re}(c)$ and $\textrm{Im}(c)$ depends on the configuration of the fixed point, i.e., the choice of $m$, $r_h$, and artificial $\Delta$. Note that when $\Delta\rightarrow0$ from the real axis, $\textrm{Im}(c)\rightarrow0$, and when $\Delta\rightarrow0$ from the imaginary axis, $\textrm{Re}(c)=\textrm{Im}(c)\rightarrow\infty$. Fig. shows that the $c$-invariant line lies between the $L\delta\kappa$-invariant line and the $L\delta\kappa$-invariant line, implying the interplay between the two roles of $\delta\kappa$, as a scale cut-off $\sim1/L$ and a gap $\sim1/\sqrt{L}$. Fig.~\ref{add} demonstrates that the $c$-invariant curve lies between the $L\delta\kappa$-invariant and $L^2\delta\kappa$-invariant curves, suggesting an interplay between the two effects of $\delta\kappa$ as a scale cutoff $\sim1/L$ and energy gap $\sim1/\sqrt{L}$. Moreover, it can be predicted that the central charge does not flow when $L^{\zeta}\delta\kappa$ ($1<\zeta<2$) remains invariant.

Returning to the unstable fixed point in the middle, we believe that it approaches $r_h=r^{*}=1$ infinitely closely as we move towards the infrared limit\footnote{ There is also a corresponding unstable fixed point $r^{*'}$ on the negative $r_h$ axis.}. At this point, both non-Bloch $\mathcal{P}$ and $\mathcal{T}$ symmetries are broken under open boundary conditions, and non-Bloch band theory is no longer applicable. From the perspective of the chain configurations shown in Fig.~\ref{core}(c), any $r_h$ perturbation will drive the system towards a different configuration, indicating that the fixed point at $r^{*}=1$ is repulsive.

We also studied two other exceptional lines, $m+2-r_h=0$ and $m-r_h=0$. Based on the slowly varying central charge flow, we claim that these belong to two other infinitely strong coupling fixed points $r^{\star}$ and $r^{\vartriangle}$, respectively, corresponding to slightly different chain configurations, and naturally their central charges differ slightly from those at $r_h=\pm\infty$ \footnote{Due to the certain similarity with $m=0$, there may also be intermediate unstable fixed points on $m+2-r_{h}=0$ and $m-r_{h}=0$, but these are not our focus, so we omit their discussion.}.

\subsection{SEPs coexisting with DPs}

We have examined the hopping parameters $v_{1,2},w_{1,2}\in\mathbb{R}$
and $a_{r}=b_{r}>0$ in the preceding sections, where real and imaginary energy disperses similarly around EPs and real band touches only at EPs. However, there is an uncommon situation when two unique types of Fermi points coexist, if the parameters are taken to be imaginary, i.e., $a_{r}=b_{r}<0$, $s\neq0$. Around EP $k_{EP}=-\pi$, the dispersion remains $\varepsilon_{k,\pm}\propto\pm\sqrt{k}$, a fully imaginary mode exists at $k_{0}=0$, and the real energy $\textrm{Re}(\epsilon_{k})\propto|k|$ surrounds it.
While the entire spectrum gap closes only at the EPs, the real spectrum gap may close when  $k_{0}=0$. Apart from the complex pairs, we found additional nontrivial pairs of real eigenvalues of the correlation matrix $0<C^{\zeta+}=1-C^{\zeta-}<1$. It resembles the non-Hermitian chiral metals' level crossing points~\cite{yi_temporal_2021,guo_entanglement_2021}. Although the total entanglement entropy is complex, the entanglement entropy calculated from the extracted additional nontrivial pairs of real eigenvalues of the correlation matrix is $S=\gamma\ln\left[\sin\left(\pi l/L\right)\right]+\textrm{const., }$ with
$\gamma\approx0.34$, which remains almost unchanged as $L$ and  $\delta\kappa$ are varied.
This undoubtedly contributes to the entanglement entropy of gapless free fermion.

\begin{figure}
\includegraphics[width=8.5cm]{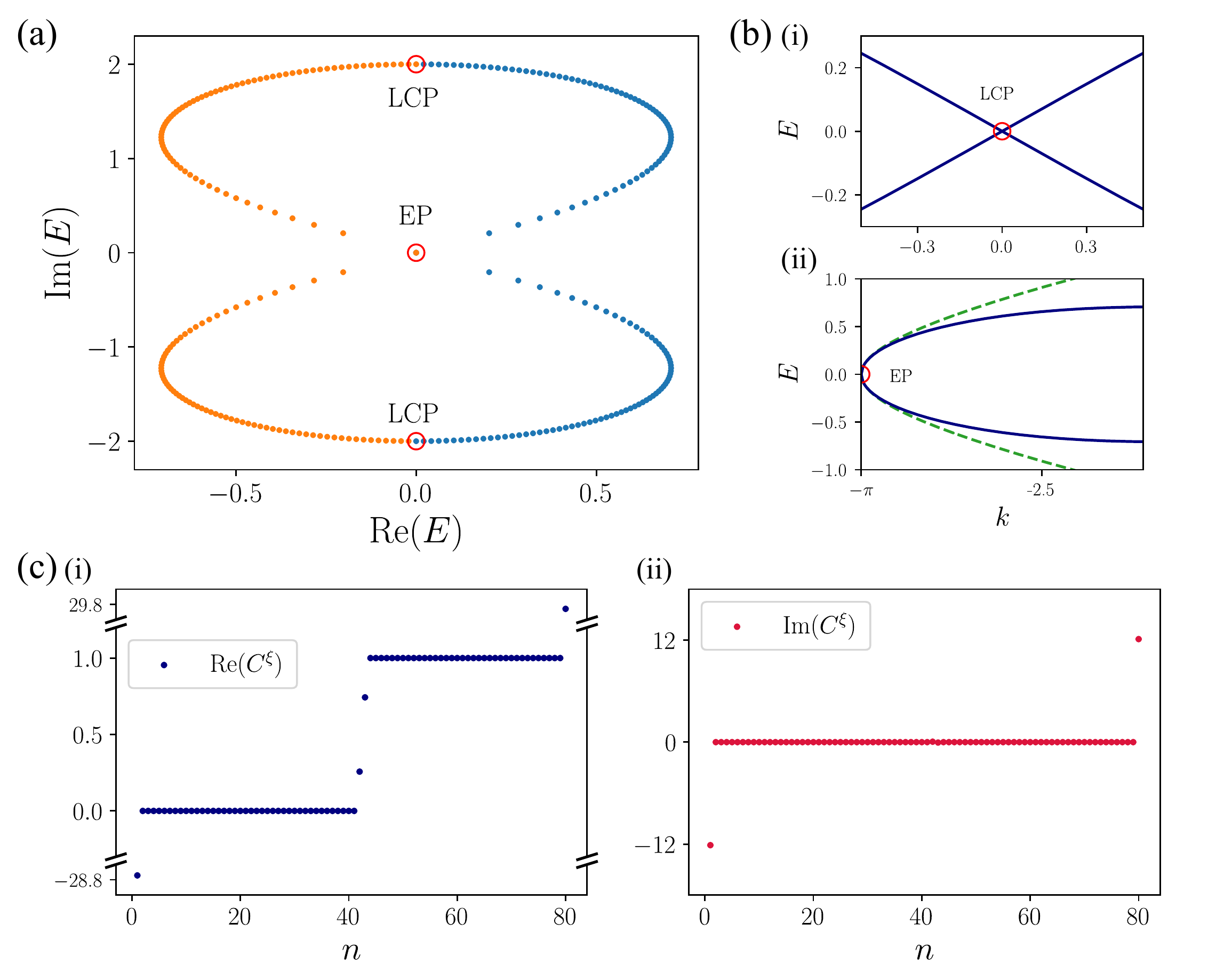}

\caption{(a) The energy spectrum of EP coexisting with level-crossing point (LCP) of the real spectrum, (b) (i) real spectra near the LCP, and (ii) spectra near EP with solid line representing the real part and dash
line representing the imaginary part, (c) (i) the real part of the eigenspectrum
of the correlation matrix in (a) with conducting states and outside exceptional
modes and (ii) the real part of the eigenspectrum of the correlation matrix in
(a) with outside exceptional modes.}
\end{figure}

\section{Entanglement spectrum and quasi-particle aspects\label{ESQ}}

Now we consider the quasi-particle properties from the aspects of the entanglement spectrum.
We add a quasi-particle (from the $\varepsilon_{k,+}$ band ) with moment $k$ above the half-filled ground
state  to the system and equally
bipartite the chain $\mathcal{S}=\mathcal{A}\cup\mathcal{B}$, $\mathcal{A}=\mathcal{B}$.
The new many-body right state is $|QPS_{R}\rangle =\psi_{R,k,+}^{\dagger}|GS_{R}\rangle $
and the left state is $|QPS_{L}\rangle =\psi_{L,k,+}^{\dagger}|GS_{L}\rangle $,
or in another form, 
\begin{equation}
\begin{array}{c}
\left|QPS_{R}\right\rangle =\frac{1}{\sqrt{2}}\left(\psi_{R,k,+}^{\mathcal{A}\dagger}+\psi_{R,k,+}^{\mathcal{B}\dagger}\right)\left|GS_{R}\right\rangle ,\\
\left|QPS_{L}\right\rangle =\frac{1}{\sqrt{2}}\left(\psi_{L,k,+}^{\mathcal{A}\dagger}+\psi_{L,k,+}^{\mathcal{B}\dagger}\right)\left|GS_{L}\right\rangle ,
\end{array}
\end{equation}
 where
$\psi_{R(L),k,+}^{\mathcal{A}(\mathcal{B})\dagger}$ is the right
(left) fermionic creating operator which creates a quasiparticle with
momentum $k$ in the A(B) region. Then the correlation matrix for A is
\begin{equation}
\label{QPE}
\begin{aligned}\mathcal{C}_{ij}^{\mathrm{Q}} & =\left\langle QPS_{L}|c_{i}^{\dagger}c_{j}|QPS_{R}\right\rangle _{i,j\in A}\\
 & =\frac{1}{2}\left(\mathcal{C}_{ij}+\left\langle GS_{L}|\psi_{L,k,+}^{\mathcal{A}}c_{i}^{\dagger}c_{j}\psi_{R,k,+}^{\mathcal{A}\dagger}|GS_{R}\right\rangle _{i,j\in A}\right)
\end{aligned}
\end{equation}

The ground state contribution is contained in the first term in $\mathcal{C}_{ij}^{\mathrm{Q}}$, whereas the quasi-particle contribution is in the second term.

For Hermitian gapped phases, the correlation matrix yields eigenvalues of 0,1 unless there is an in-gap state (topological boundary modes) that contributes a 1/2 eigenvalue, resulting in the degeneracy of the many-body entanglement spectrum as~\cite{turner,turner_band_2010,fidkowski_entanglement_2010},
\begin{equation}
    \lambda_{\{ s_{n}\} }=\prod_{n}\left[\frac{1}{2}+s_{n}\left(C^{n}-\frac{1}{2}\right)\right],\quad s_{n}=\pm1.
\end{equation}
Due to the spatial uncertainty of quasiparticle states, they act as extended bulk modes and also contribute a 1/2 in the correlation matrix spectrum when boundary modes are absent, resulting in the degeneracy of the many-body entanglement spectrum~\cite{wybo_visualizing_2021}.
While boundary modes are present, they will hybridize with the quasi-particle state, thereby lifting the degeneracy in the entanglement spectrum.

Under periodic boundary conditions, dividing the system into two parts requires two cuts, each of which hosts a topological boundary state, corresponding to the two in-gap states in the correlation matrix and the fourfold degeneracy in the entanglement spectrum. When a quasi-particle is excited, the entanglement spectrum exhibits a bulk zero mode. The two edge modes will hybridize through the extended bulk state, lifting the degeneracy in a manner analogous to the physical spectrum.

\begin{figure}
\includegraphics[width=8.5cm]{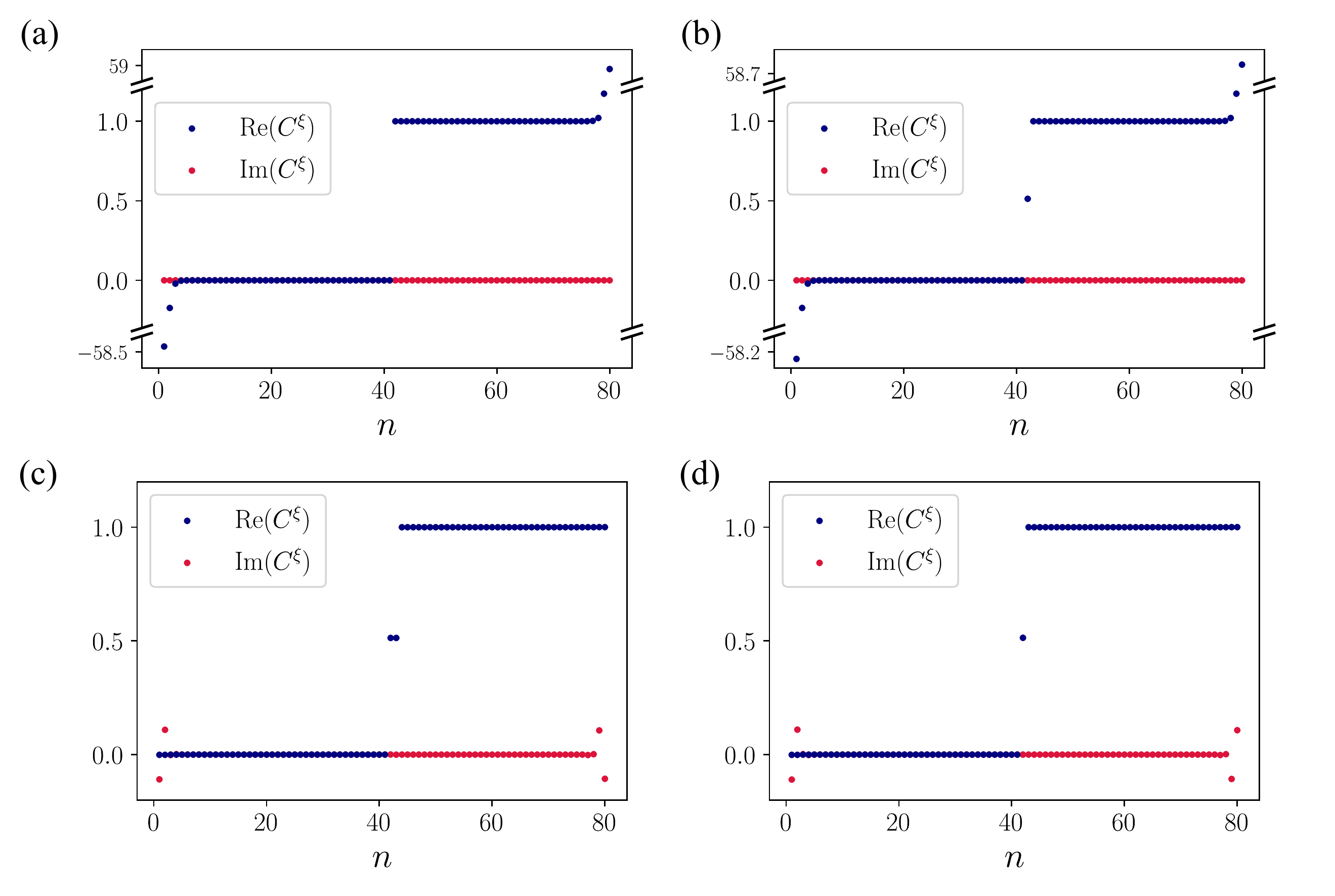}

\caption{\label{c-2es}The eigenspectrum of correlation matrix of $c=-2$ cases in PBC (a)
with exceptional mode; (b) with exceptional mode and normal quasiparticle (QP) resembling a normal zero-mode; (c) without EP but with QP, the exceptional bound modes show little hybridization with normal zero modes; (d) without EP, it behaves almost a normal metal with zero mode and exceptional bound modes. }
\end{figure}

\begin{figure}
\includegraphics[width=8.5cm]{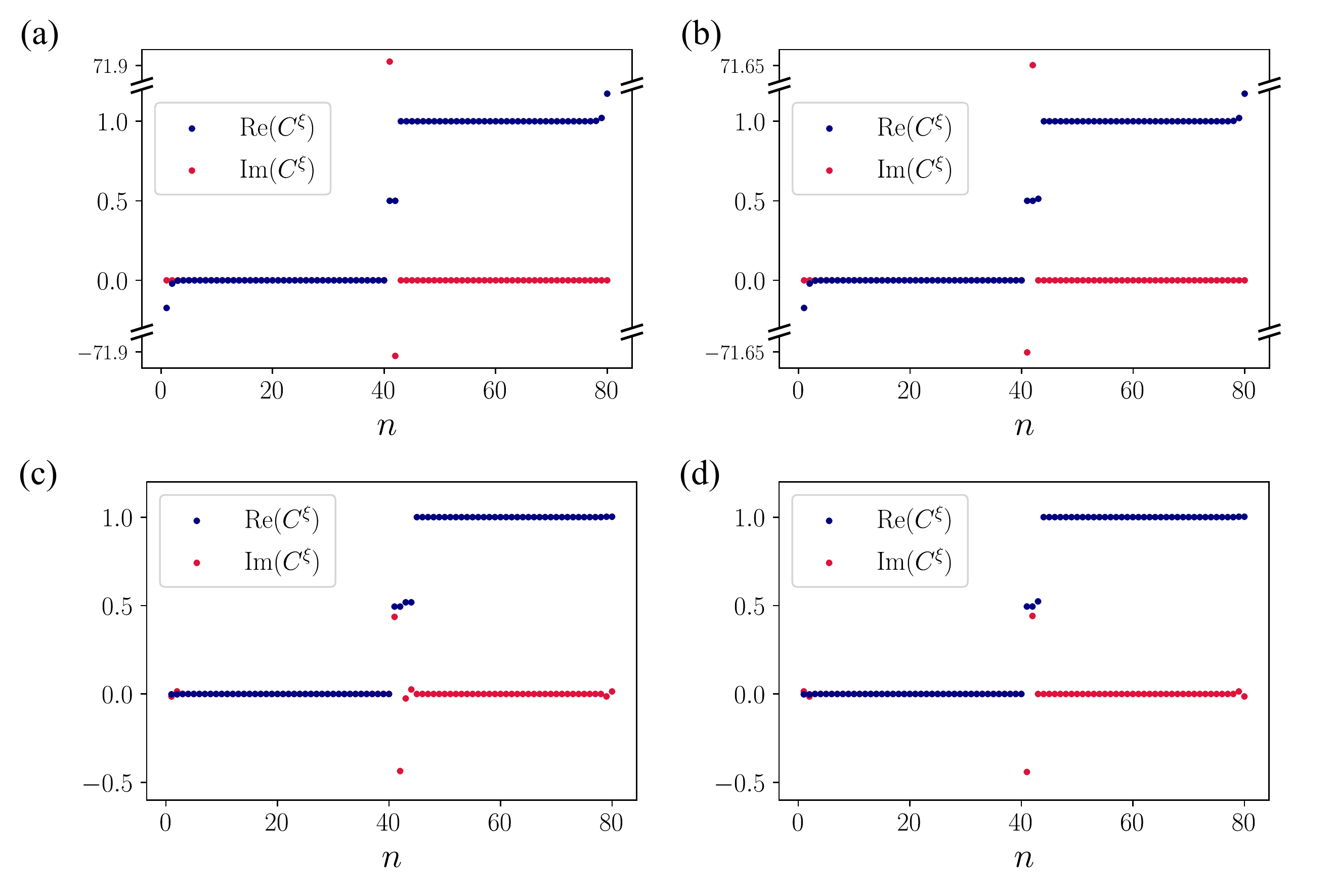}

\caption{\label{nues}The eigenspectrum of the correlation matrix of non-universal cases in Sec.~\ref{srd} (a) with exceptional mode, from real part it resembles a topological phase; (b) with exceptional mode and normal quasiparticle (QP), there are
three mid-gap states in real part; (c) without EP but with QP, the exceptional bound modes show little hybridization with normal zero mode and topological boundary modes; (d) without EP, there are three mid-gap states in real part corresponding to exceptional bound modes and two topological boundary modes and two points in the imaginary part corresponding to the two topological boundary modes. }
\end{figure}

\begin{figure}
\includegraphics[width=8.5cm]{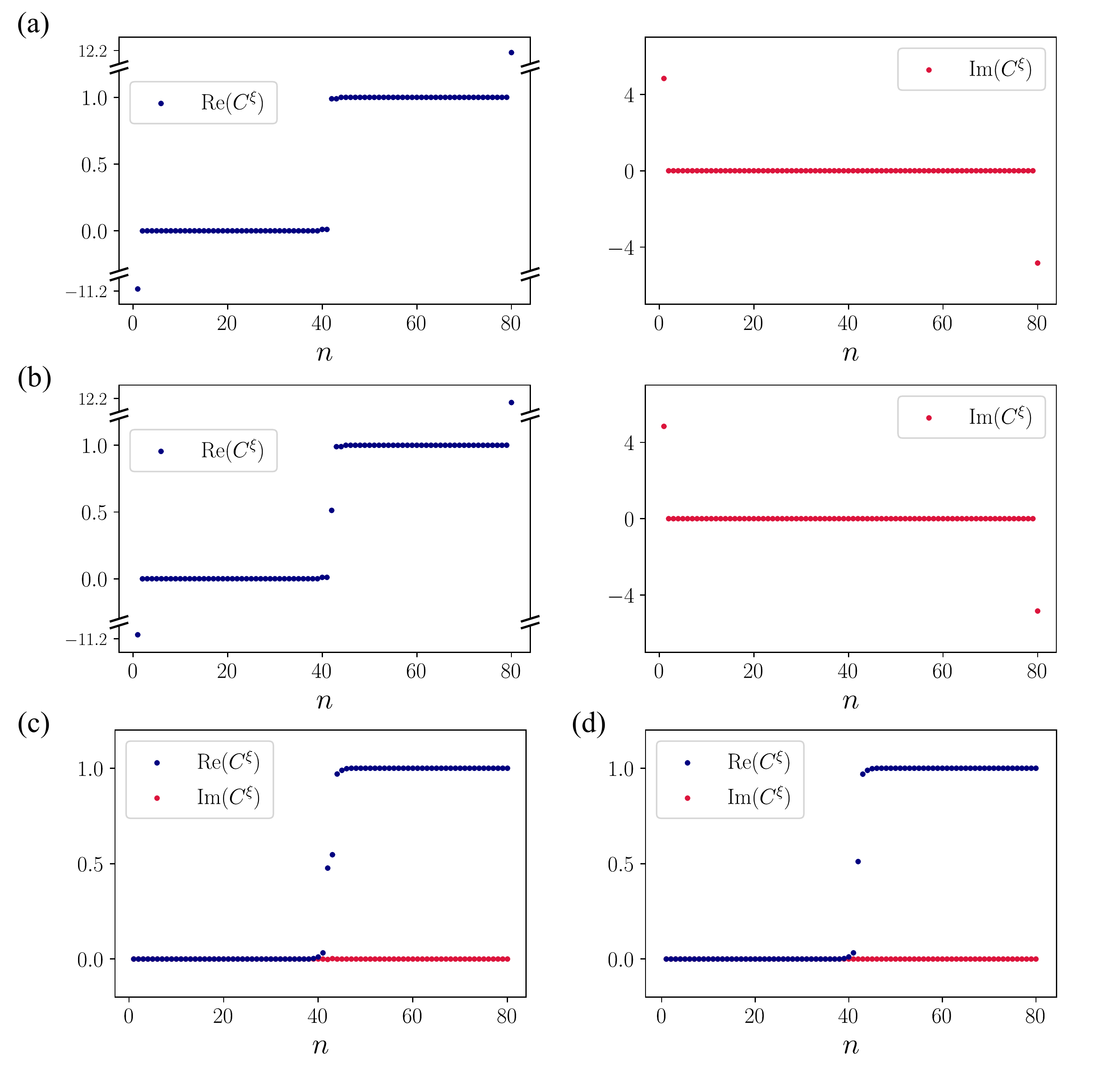}

\caption{\label{cces}The eigenspectrum of the correlation matrix of Sec.~\ref{srd} cases
in PBC (a) with exceptional mode; (b) with exceptional mode and normal quasiparticle (QP), there is a mid-gap state; (c) without EP but with QP, the non-exceptional bound modes show hybridization with normal zero mode; (d) without EP, there a bulk zero mode and several (quasi-)conducting modes due to the non-exceptional bound states.}

\end{figure}

However, in non-Hermitian scenarios, the situation is quite different. For the non-universal situation discussed in Sec.~\ref{subsec:Linear}, the entanglement spectrum does not appear to reflect the physical boundary modes. This is due to the fact that they are not necessarily of zero energy that contributes exactly 1/2.
However, the boundary modes would still manifest themselves in the $\textrm{Re}(C^{n})$ spectrum. 
For non-Hermitian gapless phases with exceptional points, there are several anomalous modes in the correlation matrix's eigenspectrum containing the information of exceptional modes, as shown in Fig.~\ref{fig:Cspectrum a}. They are highly reliant on $\delta\kappa$. Additionally, in the non-universal situation discussed in Sec.~\ref{subsec:Linear}, exceptional modes are hybridized with non-Hermitian topological modes, resulting in a significant dependence on $\delta\kappa$ of the $\textrm{Im}(C^{n})$ as shown in Fig.~\ref{fig:Cspectrum a}(d).
When a quasi-particle is excited distant from the EPs, the entanglement spectrum exhibits a normal bulk zero-mode. It is remarkable that this mode has no effect on non-Hermitian boundary modes (as illustrated in Fig.~\ref{nues}) or other non-Hermitian exceptional modes (as illustrated in Fig.~\ref{c-2es} and Fig.~\ref{cces}). This means that normal zero modes will not mix with exceptional modes or topological boundary modes. This suggests that, despite having zero energy, exceptional modes are distinct from conventional zero modes in terms of the entanglement spectrum.

Now we try to explore the entanglement aspects of other modes besides the exceptional modes. It is achieved by annihilating (or creating a corresponding quasi-hole) an exceptional mode. As a result, the entanglement spectrum undergoes a dramatic change.
The remaining exceptional bound states (non-singular states unless  taking the thermodynamic limit) dominate the entanglement spectrum. While these exceptional bound modes exhibit some of the characteristics of exceptional modes, they also exhibit certain regular metallic properties of normal zero modes.

 Exceptional bound modes result in divergent correlation functions with $L$ in the thermodynamic limit. For the $k$-linear system, the correlation matrix diverges with $L$ as $\log L$, and those exceptional bound states provide a logarithm scaling of the entanglement entropy, as in~\cite{CHLee22}. As illustrated in Fig.~\ref{c-2es} and Fig.~\ref{nues}, exceptional bound modes nearly inherit the exceptional mode's nature. While they introduce a new mid-gap state to the entanglement spectrum, it acts significantly differently from the normal zero modes. All of these anomalous entanglement modes are nearly non-hybrid with normal zero modes. This is due to the fact that exceptional bound states remain divergent.
In the case of the square root of $k$, divergent correlators with exceptional points $\delta\kappa$ now converge to a constant as $\sim\sqrt{1/L}$, which contributes no exceptional bound states. With exceptional points removed, the entanglement entropy is positive but scales non-logarithmically. the entanglement entropy curve is similar to the entanglement entropy scaling of non-realistic Lifshitz criticalities, however, their microscopic mechanism may be different. 
As seen in Fig.~\ref{cces}, the non-exceptional bound states do not contribute anomalous modes but rather conduct bulk modes that hybridize with the normal bulk mode.

\section{Analogue to non-Hermitian spin-1/2 chains\label{spin}}

 In this section, we will further examine some parallel scenarios in quantum spin models in a nutshell and try to establish a connection with the fermionic models.
We first consider a generic spin-1/2 Ising chain on a complex transverse magnetic
field
\begin{equation}
\label{complexs}
{H}_{c^{*}}=-\sum_{i=1}^{L-1}J\sigma_{i}^{x}\sigma_{i+1}^{x}-h\sum_{i=1}^{L}({\sigma}_{i}^{z}+\mathbb{I})
\end{equation}
where $J$. Parity operator, defined as
$\mathcal{P}=-\mathrm{i}\mathcal{R}^{2}=\mathrm{e}^{\frac{\mathrm{i}\pi}{2}\left(S^{z}-\mathrm{I}\right)}=\prod_{i=1}^{N}\sigma_{i}^{z}$, changes sign of $\sigma_{i}^{x},\sigma_{i}^{y}$, where rotation operator $\mathcal{R}^z=\mathrm{e}^{\frac{\mathrm{i}\pi}{4}S_{z}^{N}}=\prod_{i=1}^{N}\frac{1}{\sqrt{2}}\left(\mathbb{I}+\mathrm{i}\sigma^{z}\right)_{i}$ rotates the spins
at each site clockwise by $\pi/2$ in the $xy-$plane
and $S_{z}^{N}=\sum_{i=1}^{N}\sigma_{i}^{z}$ denotes the total spin. The time-reversal operator $\mathcal{T}$
is the usual complex conjugation, which changes the sign of $\sigma_{i}^{y}$
and the sign of the imaginary part of complex parameters.

When $h\in\mathbb{R}$, it is a transverse field Ising model (TFIM),
which can be mapped into a fermionic model via the Jordan-Wigner transformation
with periodic boundary condition $\sigma_{1}^{x}=\sigma_{L+1}^{x}$

\begin{equation}
\begin{aligned}H_{c^{*}} & =J\sum_{i=1}^{L-1}\left({c}_{i}^{\dagger}c_{i+1}+c_{i}^{\dagger}c_{i+1}^{\dagger}+\text{H.c.}\right)-2ih\sum_{i=1}^{L}\left(1-c_{i}^{\dagger}c_{i}\right)\\
 & +(-1)^{n_{F}}J\left(c_{L}^{\dagger}c_{1}+c_{L}^{\dagger}c_{1}^{\dagger}+\text{ H.c. }\right)
\end{aligned}
\end{equation}
where $(-1)^{n_{F}}=(-1)^{\sum_{i=1}^{L}c_{i}^{\dagger}c_{i}}$
denotes the fermion number parity operator, note that it is conserved,
we can fix it to be even, which corresponds to an anti-periodic boundary
condition (APBC) of the fermionic chain.
The Bogoliubov-de-Gennes Hamiltonian is 
\begin{equation}
\label{hbdg}
H_{c^{*}}=\sum_{k>0}\left(\begin{array}{cc}
c_{k}^{\dagger} & c_{-k}\end{array}\right)h(k)\left(\begin{array}{c}
c_{k}\\
c_{-k}^{\dagger}
\end{array}\right)-hL\mathbb{I}
\end{equation}
with,
\begin{equation}
h(k)=\left(\begin{array}{cc}
\epsilon_{k} & 2iJ\sin k\\
-2iJ\sin k & -\epsilon_{k}
\end{array}\right)
\end{equation}
where $\epsilon_{k}=ih+2J\cos k$. The spectrum is 
\begin{equation}
    \varepsilon_{k}=\pm2J\sqrt{1+\left(h/J\right)^{2}+2(h/J)\cos k}.
\end{equation}
Its critical point corresponds to a $c=1/2$ conformal field theory also known as Ising critical point, where dispersion is linear
around gap close point $k_{0}=-\pi$. 

\begin{figure}
\includegraphics[width=8.5cm]{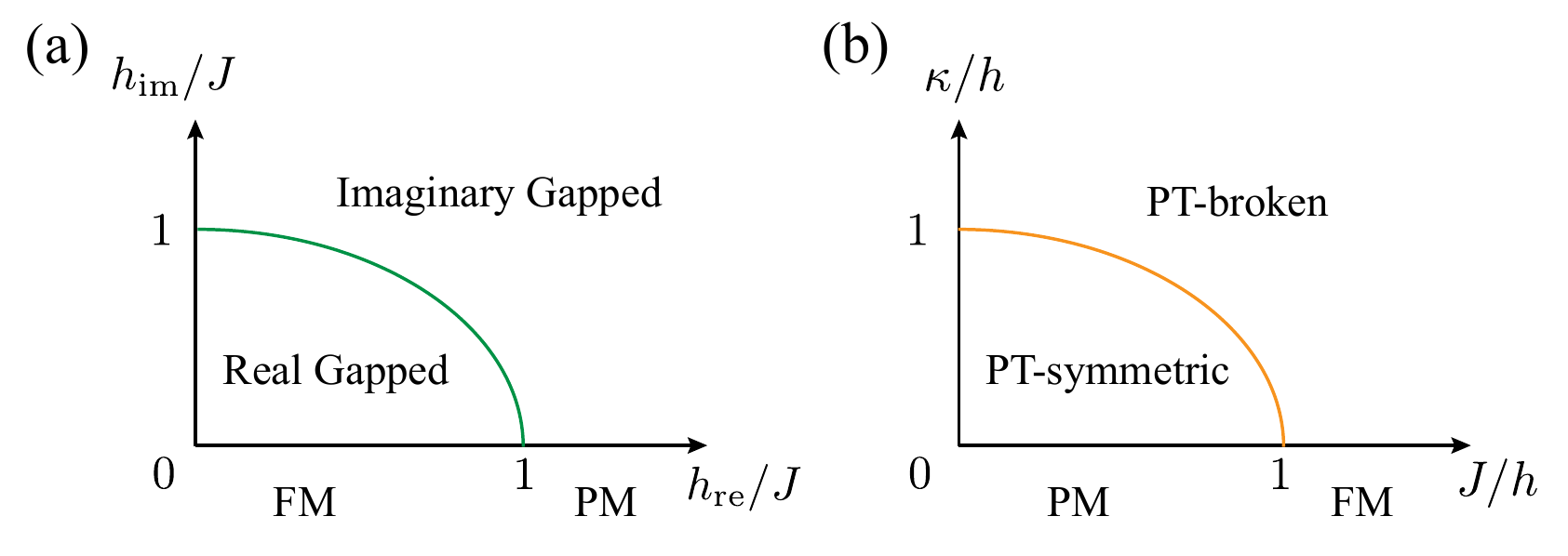}

\caption{\label{p2}Phase diagrams of two non-Hermitian spin-1/2 chain models. 
(a) Model \eqref{complexs} without $\mathcal{PT}$ symmetry. The exceptional line is depicted in green and separates the imaginary gapped phase from the real gapped phase. (b) Model \eqref{YL} with $\mathcal{PT}$ SSB. The exceptional line is colored orange and separates the $\mathcal{PT}$ symmetric and $\mathcal{PT}$ broken phases. The Ising critical point in both models is located at coordinate 1 on the transverse axis.
}

\end{figure}

When the external magnetic field
is purely imaginary, the Hamiltonian preserves $\mathcal{R}^{x}\mathcal{T}$-symmetry but no $\mathcal{PT}$-symmetry.
The Jordan-Wigner transformation is still valid, as is parity symmetry. The fermionic parity can still be fixed. Then the dispersion is
\begin{equation}
    \varepsilon_{k}=\pm2J^{x}\sqrt{1-\left(|h|/J\right)^{2}+2i|h|/J\cos k},
\end{equation}
which shares a certain similarity with Eq.~\eqref{ek}. $|h|=J$  is required for gap closing i.e,  the SEP to exist and the EPs are $k_{EP}=\pm\pi/2$. The dispersion
near SEPs is $k$-square-root. If we take a generic
complex external magnetic field, $h=h_{\textrm{re}}+ih_{\textrm{im}}$,
$h_{\textrm{re}},h_{\textrm{im}}\in\mathbb{R}$, the existence of EP requires $h_{\textrm{re}}^{2}+h_{\textrm{im}}^{2}=J^{2},$ $|h_{\textrm{re}}|,|h_{\textrm{im}}|<J$ and the EPs are $k_{EP}=\pm\arccos(-h_{\textrm{re}}/J)$. $h_{\textrm{re}}=J$ corresponds to Ising critical point. The introduction of a complex transverse field explicitly breaks $\mathcal{PT}$-symmetry, and the exceptional line divides the imaginary-energy gapped (real-energy gapless) phase and a phase with a real energy gap, as shown in Fig.~\ref{p2}(a).

The ground state of a given model is "physical" to be constructed by filling the negative (real part)  energy band. However, the presence of exceptional points (EPs) can lead to divergences, which can be mitigated by introducing a truncation on the momentum. In the case of the Bogoliubov-de-Gennes Hamiltonian, the ground state many-body entanglement entropy can be calculated using similar methods as in the preceding section. However, it should be noted that the entropy should be halved since the BdG Hamiltonian artificially doubles the degree of freedom. As a consequence, the true central charge should be half that of the corresponding fermion models. The resulting entanglement entropy exhibits a logarithmic scaling with a complex central charge and drifts with system size.

When adding a $x-$direction imaginary magnetic field on the TFIM,
\begin{equation}
\label{YL}
    H_{\text{YL}}=-\sum_{j=1}^{N}\left(h\sigma_{j}^{z}+J\sigma_{j}^{x}\sigma_{j+1}^{x}+\mathrm{i}\kappa\sigma_{j}^{x}\right),
\end{equation}
where $h,J,\kappa>0.$ This renowned model is a lattice version
of Yang-Lee edge singularity~\cite{gehlen_critical_1991,gehlen_non-hermitian_1995}, which holds $\mathcal{PT}-$symmetry. The emergence of exceptional points (EPs) at the spontaneous breaking points of $\mathcal{PT}$-symmetry separates the real spectra and bifurcated complex spectra, resulting in a transition line as shown in Fig.~\ref{p2}(b). This phase transition line connects the Ising critical point and the single-qubit  $\mathcal{PT}$-symmetry-spontaneous-breaking point, where each point corresponds to an exceptional point. It differs from the above-mentioned model \eqref{complexs}. In the thermodynamic limit, the transition line is stable and its effective field theory is related to a certain non-unitary minimal model CFT $\mathcal{M}_{5,2}$ with a central charge of $c=-22/5$. The model is highly nonlocal in fermionic representation.
The entanglement properties of the system can be studied through exact diagonalization methods, which have been previously investigated in works by Gehlen~\cite{gehlen_critical_1991,gehlen_non-hermitian_1995}. The resulting entanglement entropy leads to a positive effective central charge $c_{\textrm{eff}}=c-24\Delta_{\textrm{min}}=2/5$, with $\Delta_{\textrm{min}}=-1/5$, which is positive although non-unitary. In general, the non-unitary minimal model $\mathcal{M}{p,q}$ has an effective central charge $c{\textrm{eff}}=1-6/pq$, where $p$ and $q$ are coprime. Since $p>q\geq2$, $c_{\textrm{eff}}\geqslant0$. Therefore, these models always display non-negative entanglement entropy~\cite{Bianchini_2014}.

\begin{figure}
\centering
\includegraphics[width=8.5cm]{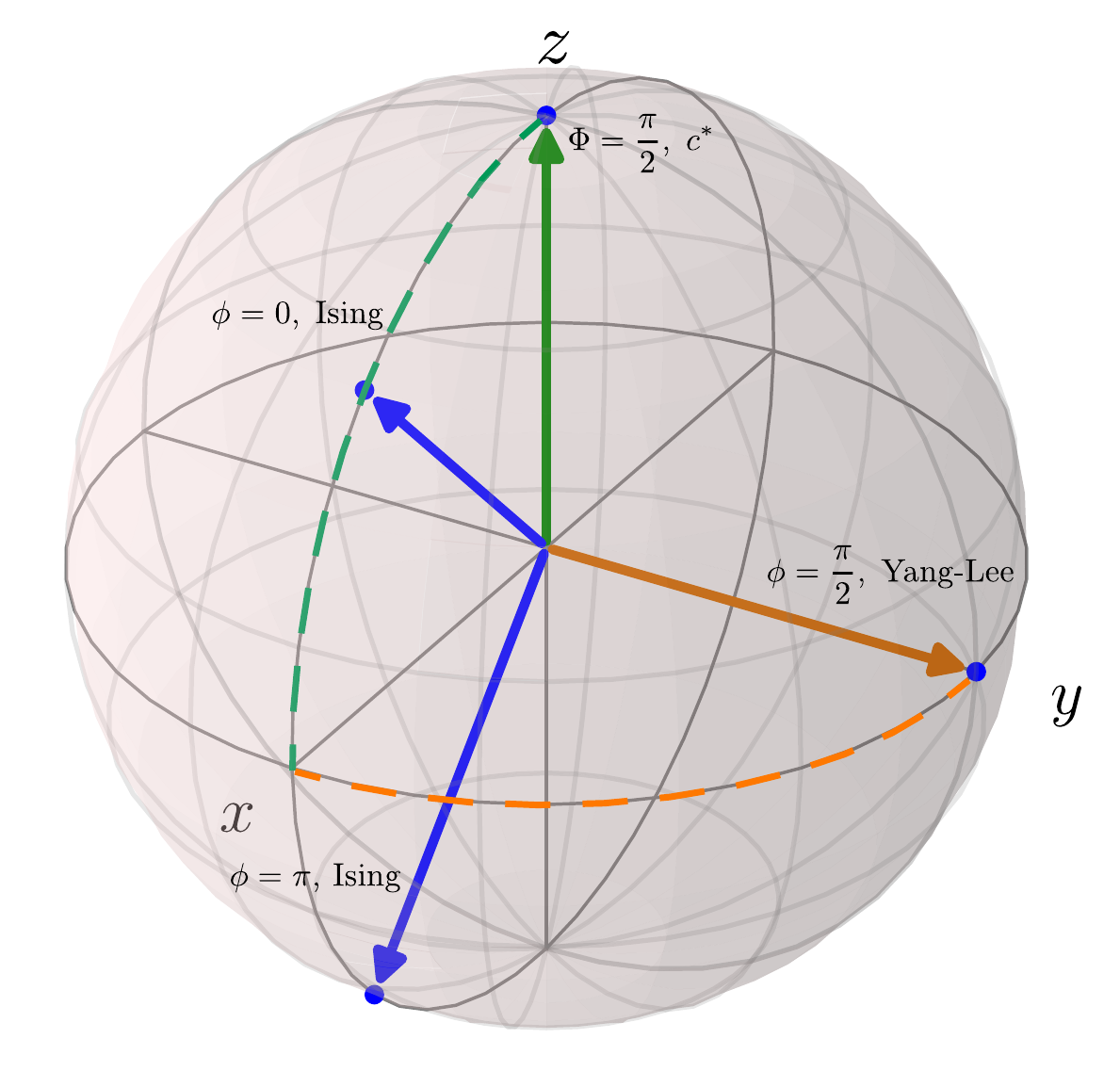}

\caption{\label{spin}Classical physical pictures of the non-Hermitian spin chain models. The (degenerate) blue vectors represent the direction of spins at the Ising critical point, while the green vector represents the direction of a spin at the exceptional point $c^{*}$ with $\Phi=\pi/2$, which corresponds to an anomalously degenerate state. The orange vector represents the spin direction of a single qubit in the Yang-Lee model \eqref{YL} with $h=\kappa, J=0$. The possible points on the green dashed line correspond to the exceptional points on the circle $h_{\textrm{re}}^{2}+h_{\textrm{im}}^{2}=J^{2}$ in Eq.~\eqref{complexs} and Fig.~\ref{p2}(a) (in the same color). On the other hand, the possible points on the orange dashed line correspond to the exceptional points on the $\mathcal{PT}$-breaking the line in Eq.~\eqref{YL} and Fig.~\ref{p2}(b) (in the same color). It should be noted that two qubits cannot describe a many-body system, and we only use this classical picture of a single spin to gain insight into the effects of the introduction of a complex magnetic field. }
\end{figure}

To gain insight into the behavior of the imaginary-magnetic-field spin chain model and obtain a "classical" physical picture, we can consider the Hamiltonian.

\begin{equation}
\label{sq}
H_s=\sigma^{z}+\exp{(i\phi)}\cdot\sigma^{x}, \phi \in [0,\pi/2],
\end{equation}

where $\sigma^{z}$ represents a spin pointing in the $z$-direction (to some extent, the chosen ferromagnetic direction), and the perturbation $\sigma^{x}$ attempts to flip the spin. As shown in Fig.~\ref{spin}, when $\phi=0$, the system is at an intermediate fixed point that corresponds to a transition between an $x$-axis-aligned state (to some extent, the chosen paramagnetic direction) and a $z$-axis-aligned state in the $xz$-plane. When $\phi=\pi/2$, the Hamiltonian lies at an EP, where only one eigenstate exists, corresponding to a $y$-axis-aligned state.  The $\phi=0$ and its mirror $\phi=\pi$ (degenerate) states can be viewed as the classical picture of a single spin near the Ising critical point. The anomalously degenerate state at $\phi=\pi/2$ is the true (single-qubit) physical picture of Eq.~\eqref{YL} with $h=\kappa, J=0$. As the system moves along the $\mathcal{PT}$-symmetry breaking line towards $\phi=\pi/2$ from the vicinity of the Ising critical point ($\phi=0$, $\phi=\pi$), the degenerate states immediately coalesce into a single state, which respects $\mathcal{PT}$ symmetry.

Then consider,
\begin{equation}
\label{dq}
H_d=\sigma^{x}\otimes\sigma^{x}+\exp{(i\Phi)}(\sigma^{z}\otimes\mathbb{I}+\mathbb{I}\otimes\sigma^{z})/2, \Phi \in [0,\pi/2].
\end{equation}
 
In the $xz$-plane, the classical picture of degenerate ground states near the Ising critical point are represented by the $\Phi=0,\pi$ points, while an anomalously degenerate state aligned with the $z$-axis (paramagnetic direction here) is represented by the $\Phi=\pi/2$ point, as expressed in Eq.~\eqref{dq} and shown in Fig.~\ref{spin}. The interval $0<\Phi<\pi/2$ emulates the exceptional line of Eq.~\eqref{complexs}. As $\Phi$ runs along this line, the degenerate states coalesce into one, and the direction of the spin gradually aligns with the $z$-axis. Notably, the $\mathcal{PT}$ symmetry is evidently broken from Fig.~\ref{spin}, as the $\mathcal{PT}$ operator corresponds to $z$-inversion.
This non-Hermitian $z$-axis alignment configuration is analogous to an ultra-violet fixed point, as demonstrated in Sec.~\ref{srd}.

Another model that is worth mentioning is the integrable
XY model subject to imaginary boundary magnetic fields~\cite{korff_ptsymmetry_2008},
\begin{equation}
    H_{g}=\frac{1}{2}\sum_{j=1}^{N}\left[\sigma_{j}^{x}\sigma_{j+1}^{x}+\sigma_{j}^{y}\sigma_{j+1}^{y}\right]+\frac{ig}{2}(\sigma_{1}^{z}-\sigma_{N+1}^{z}),
\end{equation}
which bears certain similarity with the case we study in Sec.~\ref{subsec:Linear} with $\lambda=1$ in Eq.~\eqref{lambda}. Note here
the parity operator changes signs of each $\sigma_{j}^{x},\sigma_{j}^{y}$ and
maps $\sigma_{1}^{z}$ to $\sigma_{N+1}^{z}$ and vice versa. Therefore,
the Hamiltonian is also $\mathcal{PT}-$invariant. Due to the
boundary terms, the transformed fermionic model has long-rang
hopping. The transition point occurs at $g_{c}=1$ whose effective
field theory is a nonunitary CFT with $c=-2$.

Thus, for non-Hermitian spin chain models which have exceptional points associated with spontaneous $\mathcal{PT}$-symmetry breaking, the quantum criticality of these systems are typically characterized by a non-unitary conformal field theory with a negative central charge. While those EPs that explicitly break $\mathcal{PT}$-symmetry have a fitting complex central charge from entanglement entropy, revealing the possibility of approximate conformal symmetry.
This is compatible with the non-Hermitian quadratic fermionic model results.

\section{Summary and discussion}

In this paper, we investigate the entanglement properties of non-Hermitian non-interacting fermionic models with exceptional points (EPs). The bi-orthogonal ground states of the non-Hermitian systems are generated by half-filling the real spectra and leaving a minimal cut-off in momentum space for the SEPs. We find that exceptional states have a significant effect on the measurement of entanglement.

When the dispersion around the SEPs is $k$-linear, the entanglement entropy scaling indicates that the low-energy theory of the system may be characterized by a non-unitary conformal field theory (CFT) with central charge $c=-2$. On the other hand, when the dispersion around the SEPs is $k$-square root, the complex logarithmic scaling of entanglement entropy supports the existence of underlying complex CFTs, although no apparent symmetry breaking or topological transition occurs. When the SEPs are artificially removed, all of these strange phenomena vanish, which implies the significance of SEPs to entanglement. Additionally, exceptional modes distinguish themselves from the observation of the entanglement spectrum. Similar laws are observed in non-Hermitian spin models.

Finally, it is worth noting the potential implications of these findings in higher-dimensional systems. In two dimensions, there exist non-Hermitian (semi-)metals with EPs, which can be constructed by adding non-Hermitian perturbations to a Weyl semimetal~\cite{bergholtz_exceptional_2021}. The "Fermi surface" can be either line-like (exceptional line) or point-like (Weyl node deformed to two EPs), which is expected to contribute $\alpha O(L\ln L)$ and $\beta O(L)+\gamma O(\ln L)$ respectively to entanglement entropy, where $\alpha$, $\beta$, and $\gamma$ are size and cut-off dependent complex coefficients according to a single one-dimensional EP's contribution. In three dimensions, there will be "Fermi-Seifert" surfaces in knotted non-Hermitian metals, which contain two-dimensional exceptional objects with different topology~\cite{bergholtz_exceptional_2021,knot_yang20,knots_hu21,knotted19}. The entanglement characteristics of these objects have yet to be investigated, and therefore, further research in this area is warranted.

It is anticipated that non-Hermitian systems featuring SEPs may exhibit distinct quantum criticality and universality as compared to their Hermitian counterparts. The experimental observation of such phenomena warrants further examination, given the detectability of relevant physical observables~\cite{arouca_unconventional_2020,ashida_parity-time-symmetric_2017,ashida_non-hermitian_2020}.

\section{Acknowledgments}

This work was supported by NSFC (Grants No. 11861161001), the Science, Technology and Innovation Commission of Shenzhen Municipality (No. ZDSYS20190902092905285), Guangdong Basic and Applied Basic Research Foundation under Grant No. 2020B1515120100, Shenzhen-Hong Kong Cooperation Zone for Technology and Innovation (Grant No.HZQB-KCZYB-2020050), and Center for Computational Science and Engineering at Southern University of Science and Technology.

\bibliography{exceptional}

\end{document}